\documentclass[twocolumn]{aastex6}
\usepackage{natbib}
\usepackage{amsmath,amssymb}
\usepackage{wasysym}
\usepackage{graphicx,epsf}
\def\ve#1{\boldsymbol{#1}}
\begin{document}

\title{A weak lensing view of the  downsizing of star-forming galaxies}\thanks{Based on data collected at Subaru Telescope, which is operated by the National Astronomical Observatory of Japan.}
\author{Yousuke Utsumi\altaffilmark{1}}\email{youtsumi@hiroshima-u.ac.jp}
\author{Margaret J. Geller\altaffilmark{2}}
\author{Ian P. Dell'Antonio\altaffilmark{3}}
\author{Yukiko Kamata\altaffilmark{4}}
\author{Satoshi Kawanomoto\altaffilmark{4}}
\author{Michitaro Koike\altaffilmark{4}}
\author{Yutaka Komiyama\altaffilmark{4,5}}
\author{Shintaro Koshida\altaffilmark{6}}
\author{Sogo Mineo\altaffilmark{4}}
\author{Satoshi Miyazaki\altaffilmark{4,5}}
\author{Jyunya Sakurai\altaffilmark{4,5}}
\author{Philip J. Tait\altaffilmark{6}}
\author{Tsuyoshi Terai\altaffilmark{6}}
\author{Daigo Tomono\altaffilmark{6}}
\author{Tomonori Usuda\altaffilmark{4,5}}
\author{Yoshihiko Yamada\altaffilmark{4}}
\author{Harus J. Zahid\altaffilmark{2}}

\altaffiltext{1}{Hiroshima Astrophysical Science Center, Hiroshima University, 1-3-1 Kagamiyama, Higashi-Hiroshima, Hiroshima, 739-8526, Japan}
\altaffiltext{2}{Smithsonian Astrophysical Observatory, 60 Garden Street, Cambridge, MA 02138, USA}
\altaffiltext{3}{Department of Physics, Brown University, Box 1843, Providence, RI 02912, USA}
\altaffiltext{4}{National Astronomical Observatory of Japan, 2-21-1 Osawa, Mitaka, Tokyo 181-8588, Japan}
\altaffiltext{5}{Department of Astronomical Science, The Graduate University for Advanced Studies (SOKENDAI), 2-21-1 Osawa, Mitaka, Tokyo 181-8588, Japan}
\altaffiltext{6}{Subaru Telescope, National Astronomical Observatory of Japan, 650 North A'ohoku Place, Hilo, HI 96720, USA}

\begin{abstract}
We describe a weak lensing view of the downsizing of star forming galaxies based on  cross correlating a weak lensing ($\kappa$) map with a predicted map constructed from a redshift survey.
Moderately deep and high resolution images with Subaru/Hyper Suprime-Cam 
covering the $4{\rm deg}^2$ DLS F2 field provide a $\kappa$ map with 1 arcmin resolution.
A dense complete redshift survey of the F2 field including 12,705 galaxies with $R\leq20.6$ is the basis for construction of  the predicted map.
The  zero-lag cross-correlation between the $\kappa$ and predicted maps is significant at the 30$\sigma$ level.
The width of the cross-correlation peak is comparable with the angular scale of rich clusters at $z\sim0.3$, the median depth of the redshift survey.
Slices of the predicted map in $\delta{z} = 0.05$ redshift bins enable exploration of the impact of structure as a function of redshift.
The zero-lag normalised cross-correlation has significant local maxima at redshifts coinciding with  known massive X-ray  clusters.
Even in slices where there are no known massive clusters, there is  significant signal in the cross-correlation
originating from lower mass groups that trace the large-scale of the universe.
Spectroscopic $D_n4000$ measurements enable division of the sample into star-forming and quiescent populations. In regions surrounding massive clusters of galaxies,
the significance of the cross-correlation with maps based on  star-forming galaxies
increases with redshift from 5$\sigma$ at $z = 0.3$ to 7$\sigma$ at $z = 0.5$; the fractional
contribution of the star-forming population to the total cross-correlation signal also increases with redshift. 
This weak lensing view is consistent with the downsizing picture of galaxy evolution established from  other independent studies.
\end{abstract}


\section{Introduction}
Weak lensing has come of age as a tool for mapping the large-scale matter distribution in the
universe. Wide-field cameras on 4-8m telescopes have enabled deep, high quality imaging of large regions of the sky \citep{2012MNRAS.427..146H,2016MNRAS.tmp..827J,2012SPIE.8446E..0ZM}. The resulting weak lensing maps trace the projected foreground matter distribution. Foreground galaxies in the appropriate redshift window also trace this matter distribution. Thus comparisons of lensing maps with maps derived from the galaxy distribution are potentially powerful cosmographic probes. 

\cite{1998astro.ph..9268K}  first cross-correlated a weak lensing map of a supercluster with a predicted map based on the foreground distribution of early-type galaxies.
They demonstrate that the salient features in the weak lensing map correspond very well with the foreground structure traced by early-type galaxies. \cite{2001ApJ...556..601W} extended this approach to a 0.5$\times 0.5$ deg$^2$ field. In this more general region, their cross-correlation approach demonstrates the correspondence between
the projected surface mass density revealed by the lensing map and that predicted from the foreground distribution of early type galaxies.  

More recently \cite{2013MNRAS.433.3373V} and \cite{2015PhRvD..92b2006V} explored weak lensing maps covering more than 100 square degrees. In both studies, the comparison maps derived from the foreground galaxy distribution are based on photometric redshifts. \cite{2013MNRAS.433.3373V} derive a projected mass density from the galaxy distribution and \cite{2015PhRvD..92b2006V} derive a projected number density of galaxies. In both studies the cross-correlation between the weak lensing map and the map predicted based on the foreground galaxies is significant and
$\gtrsim 7-10\sigma$.  All of these studies highlight the promise of these combined approaches for understanding the matter distribution in the universe and the way galaxies of various types trace it.

\cite{2005ApJ...635L.125G} made an early exploration of the use of a dense foreground redshift survey in the construction of a map for comparison with weak lensing. In contrast with earlier work, they use the velocity dispersion among galaxies as a function of position as a proxy for the projected mass density. Although their resulting maps have fairly low spatial resolution, they detected a cross-correlation with the Deep Lens Survey lensing map of the F2 region \citep{2002SPIE.4836...73W,2006ApJ...643..128W}
at the $~7\sigma$ level. \cite{2005ApJ...635L.125G} detected the cross-correlation not only for the
full redshift range effectively imaged by the weak lensing map but also for redshift slices.

Here we once again explore the matter distribution in the F2 region of the Deep Lens 
Survey. We derive a weak lensing map from Hyper Suprime-Cam (HSC) i-band observations in 0.5
to 0.7 arcsec  seeing (in contrast with the 0.9 arcsec  seeing for the DLS map). We compare the weak lensing map with a dense, deep foreground redshift survey complete to $R = 20.6$  \cite[hereafter G14]{2014ApJS..213...35G} in contrast with \cite{2005ApJ...635L.125G} who used a redshift survey complete to $R\sim19.7$. The weak lensing map and predicted maps derived from the redshifts
enables a fresh assessment of the way galaxies trace the large-scale matter distribution at redshifts 0.1 $ < z < 0.7$ in this four square degree patch of the universe.

We use cross-correlation of the  weak lensing and redshift survey predicted maps as a tool
for investigating the morphological correspondence (or lack of it) among the maps. The redshift survey enables investigation of the cross-correlation in narrow redshift slices
that elucidate the contribution to the weak lensing signal by rich clusters identified with other techniques
\citep[e.g. the X-ray;][]{2014ApJ...786..125S}. 

The redshift survey also sets the stage for a weak lensing view of the role of massive star-forming galaxies as tracers of the matter distribution. G14 and \cite{2016ApJS..224...11G} demonstrate the increasing fraction of massive star-forming galaxies with redshift
in their surveys of the DLS fields F2 and F1, respectively. This behavior corresponds to the well-studied down-sizing of star-forming galaxies first observed by \cite{1996AJ....112..839C}. Based on the spectroscopy, we construct predicted maps for quiescent and star-forming galaxies and then cross-correlate these maps with the HSC weak lensing map. Remarkably the 
significance of the cross-correlation signal  in regions around massive clusters increases with redshift, complementing other views of the way the star-forming galaxies trace the matter distribution in and around rich clusters at increasing redshift.

Several investigators cross-correlate  sets of weak lensing  peaks with a galaxy catalog \citep{2016MNRAS.456.2806B, 2016MNRAS.tmp.1360C,2014MNRAS.442.2534S}. This approach complements but differs from cross-correlating the entire maps. Cross-correlating the full maps takes all of the structure into account including, for example, low density regions projected on the sky.

We describe the maps derived from the HSC data and from the redshift survey in Section \ref{sec:maps}.
Sections \ref{shearcatalog} and \ref{weaklensmap} discuss the construction and testing of the weak lensing map and Section \ref{predictedmap} describes the construction of the predicted map from the redshift survey.
We cross-correlate the maps in Section \ref{sec:crosscorrelation}.
Section \ref{sec:zslices} focuses on the cross-correlation in redshift slices and Section \ref{starforming} highlights the power of weak lensing for investigating the role of star-forming galaxies as tracers of the matter distribution. We discuss the limitations and implications of these results in Section \ref{sec:discussion} and we conclude in Section \ref{sec:conclusion}.

Unless otherwise stated, we the WMAP9 cosmological parameters: $\Omega_m = 0.286$, $\Omega_\Lambda = 0.713$ and $h = 0.693$ \citep{2013ApJS..208...19H}.

\section{Maps}\label{sec:maps}
\subsection{The Data}
\subsubsection{HSC Imaging}

\begin{figure}[htbp] 
   \centering
	   \includegraphics[bb=0 0 580 435, width=4in]{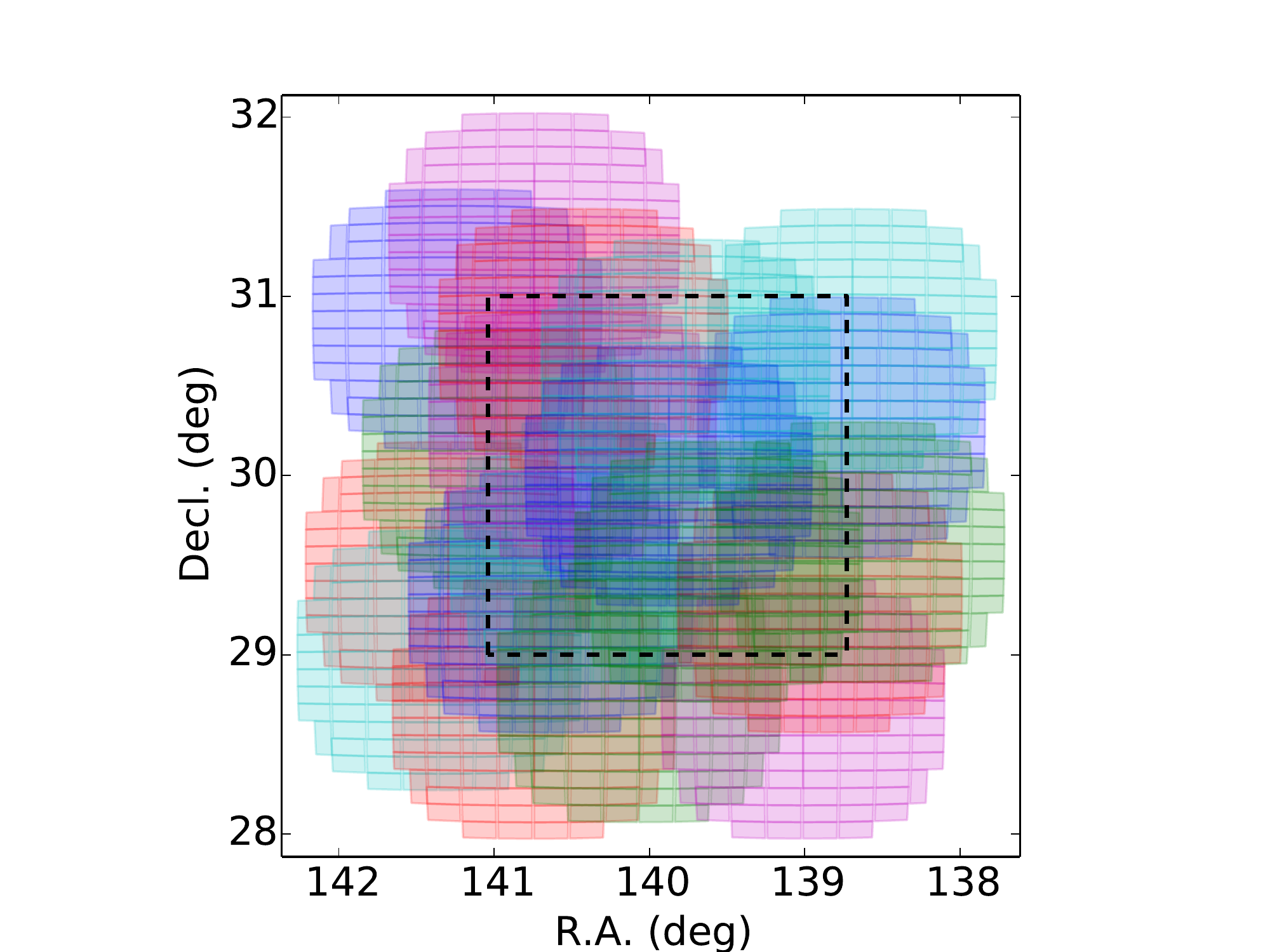}
   \caption{Pointing map for the HSC data. Each round colored region represents a distinct pointing. The dashed box show the original DLS F2 footprint.}
   \label{fig:patches}
\end{figure}

We imaged a $2\times2~{\rm deg}^2$  region covering the Deep Lens Survey (DLS) \citep{2002SPIE.4836...73W} field F2 centered at $(\alpha, \delta)=(9^{h}18^{m}0^{s},  +30^{\circ}00'00")$ with Subaru / Hyper Suprime-Cam  \citep[HSC;][]{2012SPIE.8446E..0ZM}.
During an HSC engineering run on 2014/11/30 we imaged the field in the HSC $i$-band (hereafter $i$-band) with a 240 second exposure  for each pointing.
To reach a uniform depth across the approximately 4 square degree field, the pointings extend beyond the 
original boundaries of the F2 field (Figure \ref{fig:patches}). The typical seeing for the HSC imaging 
is in the range 0.5 to 0.7 arcsec.

We verified the reliability of the HSC auto guiding system \citep{2008SPIE.7014E..4WM,2012SPIE.8446E..62U} by observing a blank field.
If the stellar images were elongated as a result of failure to acquire an appropriate guide star, we discarded and then repeated the exposure.

The {\it hscPipe} system is the standard pipeline developed for reduction and analysis of the HSC Subaru Strategic Program (SSP) \footnote{http://hsc.mtk.nao.ac.jp/}.
We ran version 3.10.2 to reduce the F2 data. 
{\it hscPipe} de-biases the images by evaluating a bias level from the overscan region. Trimming, 
flat-fielding, cosmic ray rejection, and an astrometric solution based on a 5th order  polynomial fit to the SDSS DR9 catalog follow the initial de-biasing.
After reducing all of the individual chips, {\it hscPipe} improves the mosaicking solution by fitting a 9th order polynomial to the
combination of all of the chips in a grid of  overlapping  $1.7\times1.7$ deg$^2$ square degree subregions of the F2 imaging data. In the HSC reduction system these subregions are called tracts.

With the resulting accurate CCD positions and flux scaling, {\it hscPipe} warps the individual images and stacks them by
taking the arithmetic mean and rejecting deviant pixels. 
{\it hscPipe} performs photometry at each stage of the reduction, but we only use positions and apparent magnitudes of the galaxies from the final stacked image. However, at each galaxy position we measure the shape of each source from the individual exposures.

\subsubsection {The SHELS Redshift Survey of the F2 Field}\label{SHELSDESCRIPTION}

G14 used the Hectospec wide-field multi-object spectrograph
\citep{2005PASP..117.1411F} mounted on the MMT to carry out a complete redshift survey of the F2 field called SHELS. DLS imaging provided the photometric catalog for the redshift survey \citep{2002SPIE.4836...73W}.

The redshift survey is 95\% complete to a limit $R = 20.6$ where the magnitudes are extrapolated Kron-Cousins $R$-band total magnitudes. The total area of the DLS F2 field is
4.19 deg$^2$. In approximately 5\% of the area, the photometry is corrupted by nearby bright stars. G14 mask this area and their redshift survey thus covers 3.98 deg$^2$. The masked area is irrelevant for comparison with the weak lensing map we construct from the HSC data because we mask the same area in the lensing map construction.

G14 discuss the details of the spectroscopy. The total number of galaxies in the complete survey is 12,705. The 708 galaxy candidates brighter than the magnitude limit and without a redshift
are mostly near the corners and edges of the field (see Figure 4 of G14). Their effect on the analysis here is negligible and we make no correction to account for them. 
\begin{figure}[htbp] 
   \centering
	   \includegraphics[bb=0 0 580 435, width=4in]{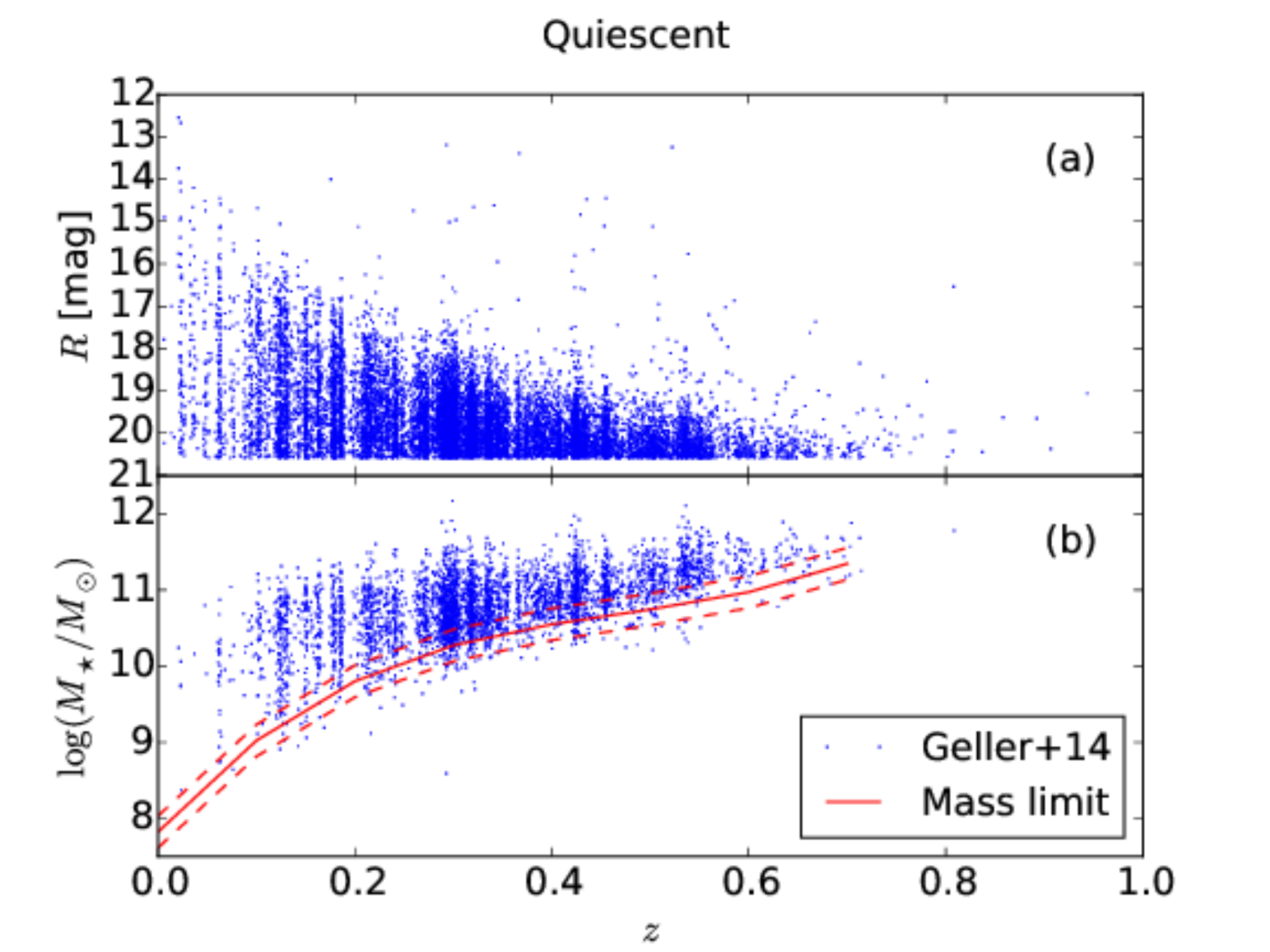}
   	   \includegraphics[bb=0 0 580 435, width=4in]{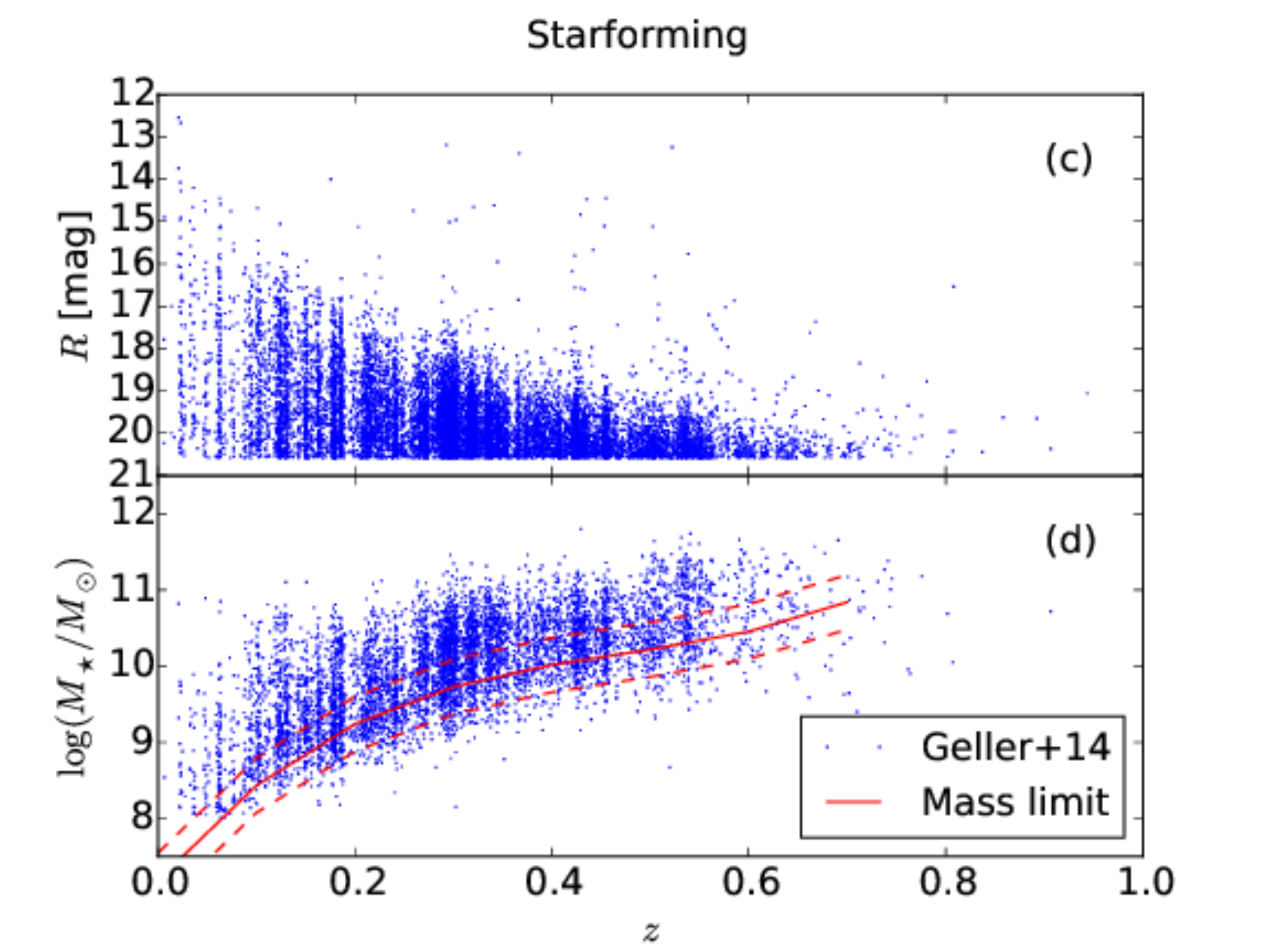}
   \caption{Apparent magnitude distributions for galaxies in the F2 redshift survey (panels (a) and (c)). Panel (a) shows the distribution for quiescent objects and panel (c) shows the distribution for star-forming objects classified on the basis of $D_n$4000.
   Panels (b) and (d) show the distribution of stellar masses as a function of redshift for the quiescent and star-forming populations, respectively. The solid curve is a fit to the effective stellar mass limit as as a function of redshift and the dotted lines indicate the 1$\sigma$ scatter in this limit.  }
   \label{fig:specsamples}
\end{figure}
For each galaxy in the redshift survey, G14 include a $D_n$4000 and a stellar mass.
They determined $D_n$4000 as the ratio of flux (in $f_\nu$ units) in the $4000-4100 {\rm \AA}$ and $3850-3950{\rm \AA}$ bands \citep{1999ApJ...527...54B}. For the SHELS data, the typical error
in the $D_n$4000 (based on 1468 repeat measurements) is 0.045 times the value of the index.

The SHELS values for $D_n$4000 are also in essential agreement with those derived for 
overlapping SDSS galaxies.
\cite{2010AJ....139.1857W} demonstrate the $D_n$4000 is  useful for segregating
quiescent and star-forming galaxies. Following \citet{2010AJ....139.1857W} we use $D_n$4000 = 1.5
as the dividing line between the two populations.
Figure \ref{fig:specsamples} shows the distribution of apparent magnitudes as a function of redshift for both quiescent ($D_n 4000\geq 1.5$; upper panels) and star-forming galaxies
($D_n4000 < 1.5$; lower panels).

Stellar masses for the SHELS galaxies are also available in G14.
These stellar masses are derived by applying the {\it LePhare} code \citep{1999MNRAS.310..540A,2006A&A...457..841I} to the SDSS five band photometry for galaxies in the SHELS F2 survey.
G14 note that there is a systematic offset between these values and the stellar masses derived by the MPA/JHU group.
After correction for this offset, the scatter between the G14 and MPA/JHU estimates is 0.17 dex, consistent with the error in the two techniques.
The offset is irrelevant for our analysis and we simply take the stellar masses directly from G14.
The upper and lower panels of Figure \ref{fig:specsamples} show the distribution of stellar masses for the quiescent and star-forming galaxies, respectively.

\begin{figure}[htbp] 
   \centering
	   \includegraphics[bb=0 0 580 435, width=4in]{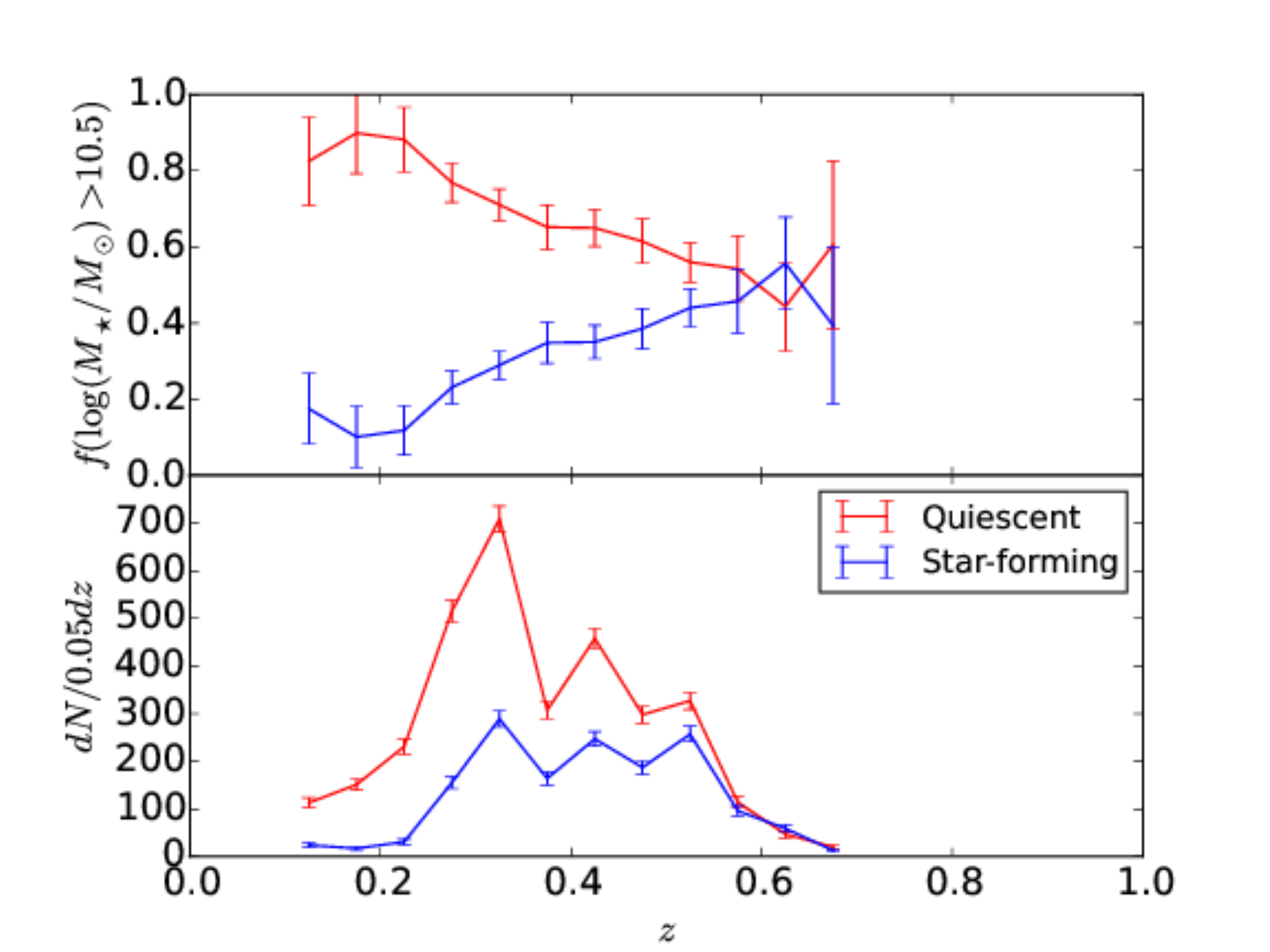}
   \caption{ (Top) Fractions and (Bottom) differential number counts of quiescent (red)  and star forming (blue) galaxies for stellar masses $M>10^{10.5}M_{\odot}$ in the SHELS survey as a function of redshift.  As expected in the downsizing picture, most massive galaxies at $z = 0.2$ are quiescent. At $z\sim 0.6$, $\sim 50$\% of these massive galaxies are star-forming.}
   \label{fig:fraction}
\end{figure}

G14 (Figure 13) shows distributions of $D_n4000$ as a function of redshift and stellar mass.
Their display reveals the well-known down-sizing of star-forming galaxies.
In other words, the fraction of massive star-forming galaxies increases with redshift.
Figure \ref{fig:fraction} summarizes this result and shows how the fraction of star-forming galaxies with $M_{*} \geq 10^{10.5}M_{\odot}$ increases with redshift in the SHELS F2 survey region.  The lower panel of Figure \ref{fig:fraction} shows the number of massive quiescent (red) and star-forming (blue) galaxies as a function of redshift and demonstrates the impact of large-scale structure. 

\subsection {Measuring the Weak Lensing Shear}\label{shearcatalog}

We follow the procedure described in \citet{2015ApJ...807...22M} to construct a shear catalog.
Here we summarize the procedure.
We use {\it lensfit} \citep{2007MNRAS.382..315M, 2008MNRAS.390..149K}  to measure galaxy shapes and to derive a reduced shear $\ve{g}=\ve{\gamma}/(1-\kappa)$ where $\kappa$ is the lensing convergence, and $\ve{\gamma}$ is the shear.

{\it lensfit} is a Bayesian shape measurement code that constructs individual galaxy models.
This reduction system can measure shapes not only from the stacked image but also from each exposure.
To apply {\it lensfit} we first  construct a galaxy catalog  and a star catalog.  
We used "extendedness"  provided by \emph{hscPipe} and calculated from a combination of second-order flux moments to separate stars from galaxies.
We reject an object if it has a bad pixel flag.
Stars have "extendedness=0.0" and galaxies have "extendedness$\neq$0.0".
We also apply a magnitude cut $20.0<i<23.0$ to the star catalog, and a cut of $23<i<25.5$ to the galaxy catalog  to select background lensed objects.

We use the star catalog to make the PSF anisotropy correction. There are
44,446 stars stored in the star catalog.

\begin{figure}[htbp] 
   \centering
	   \includegraphics[bb=0 0 580 435, width=4in]{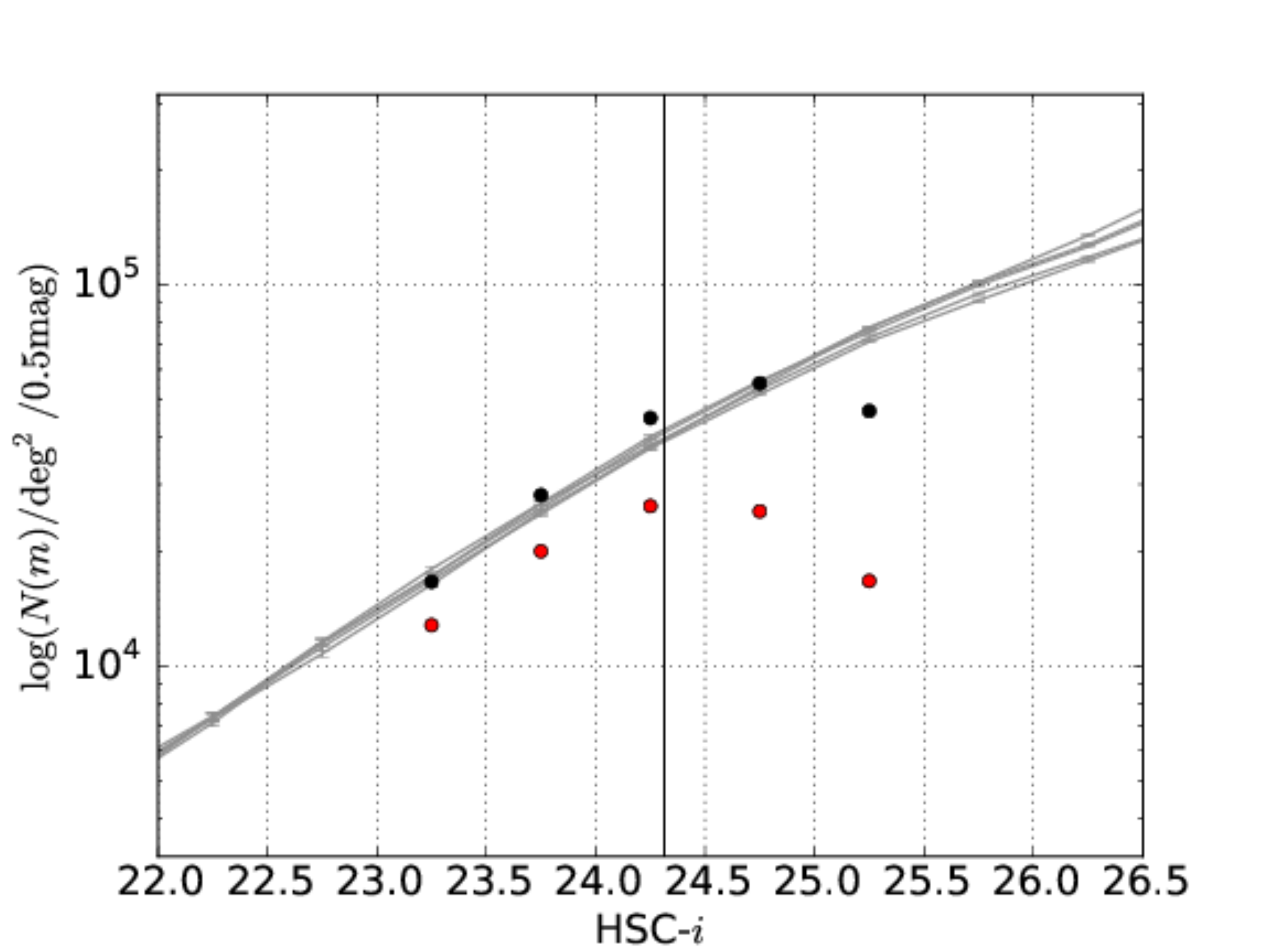}
   \caption{Number count of galaxies in the HSC images as a function of $i$-band apparent magnitude.
Black dots show the counts for the total initial catalog of galaxies detected (2,363,558 objects); red dots show the source galaxies (1,248,649 objects) with measured ellipticities.
The total counts decline for $i\gtrsim25$ where the catalog becomes incomplete;
the corresponding counts of source galaxies turn over at a brighter apparent magnitude because higher signal-to-noise is necessary to measure the shape.
   The vertical solid line shows the mean apparent magnitude for the source galaxies.
   The five curves show deep number counts based on Suprime-Cam imaging \citep{2008ApJS..176....1F}.
   The agreement between the HSC and Suprime-Cam counts is excellent. }
   \label{fig:numbercount}
\end{figure}

We apply {\it lensfit} to our initial
catalog of  2,363,558 galaxies following the procedure described in \citet{2015ApJ...807...22M}. 
We fit 1,248,649 galaxies successfully (fitclass = 0  \& weight $>$ 0). 
We reject galaxy candidates with fitclass $\neq$ 0. The parameter
fitclass $\neq$ 0  if the fitting procedure in lensfit finds that
(1) the reduced $\chi^2 \geq 1.4$ for the fit 
(2) the likelihood surface is badly behaved, for example the maximum in the surface is outside the nominal position error,
(3) the source is blended,
(4) there are actually no data at the specified coordinate or (5) the object is probably a star.
At magnitudes $i \gtrsim 24$, a combination of poor signal-to-noise and
resolution prevents detection and classification of all of the possible galaxy sources;
thus the total counts of galaxies shown in Figure \ref{fig:numbercount} do not continue to increase at fainter magnitudes.
Our total counts (black points in Figure \ref{fig:numbercount}) in the interval $i = 23 -24.5$  agree with the counts from \citet{2008ApJS..176....1F} for the Suprime-Cam Subaru/XMM-Newton Deep Survey (SXDS).

The mean magnitude for the galaxies included in our shear catalog is $\langle i\rangle=24.3$ (vertical line in Figure \ref{fig:numbercount}).
The corresponding  mean source galaxy redshift is $\langle z \rangle =1.12$
according to the relation derived from the  COSMOS/HST/ACS photometric redshift catalog \citep{2010A&A...516A..63S}. The average lensed source number density is 34 ${\rm arcmin}^{-2}$.

The {\it lensfit} software was developed for CFHT lensing measurements. The CFHT
observations are much shallower than the HSC imaging. Thus  potential
additive and  multiplicative biases in the shear measurement
must be calibrated \citep{2013MNRAS.429.2858M}.

To check our shear catalog, we compute weighted median statistics for the reduced shear over the entire observed field. The shear components, $g_1$ and $g_2$, are both consistent with 0
to within the statistical error:
$\langle g_1, g_2\rangle = (-0.00058, -0.00023) \pm (0.0006, 0.0006)$. This result demonstrates that any
additive bias in the shear measurement is below the level of the statistical error and is thus
negligible for our purposes.  We do not explicitly examine the impact of PSF residuals or position-dependent additive biases as required to measure cosmic shear; treating these effects is unnecessary for our application.

To avoid the impact of multiplicative bias, we  focus on normalized cross-correlations
to evaluate the relationship between the lensing map  and the foreground structure in the galaxy distribution. Because we specialize to cross-correlation measures exclusively, we can avoid
calibrating any multiplicative bias in the lensing map.

\subsection{The weak lensing map}\label{weaklensmap}

The weak lensing map ($\kappa$) is the convergence map derived from the weighted tangential shear
\begin{eqnarray}
	\kappa(\ve{\theta}) = \int d^2 \phi \gamma_t (\ve{\phi};\ve{\theta})  Q(|\ve{\phi}|)\label{eq:kappa}
\end{eqnarray}
where $\gamma_t(\ve{\phi};\ve{\theta})$ is the tangential component of the shear at position $\ve{\phi}$ relative to the point $\ve{\theta}$,
and $Q$ is the weight function
\begin{eqnarray}
	Q(\theta) = \frac{1}{\pi \theta^2}\left[ 1 - \left(1+\frac{\theta^2}{\theta_G^2}\right)\exp\left( -\frac{\theta^2}{\theta_G^2}\right)\right]
\end{eqnarray}
for $\theta<\theta_o$ and $Q=0$ elsewhere. Here $\theta_G$ is a smoothing scale and $\theta_o$ is the truncation radius for the Gaussian smoothing. In the weak lensing limit, we assume $g\simeq\gamma$.

Weak lensing effectively images the projected mass distribution traced by foreground galaxies. For source galaxies at $\langle z\rangle\sim1.12$,
the lensing sensitivity  peaks at $z\sim0.4$ but is broadly sensitive to the foreground structure in the redshift range $0.1<z<0.7$ \citep[e.g.][Figure 14]{1998astro.ph..9268K}.
Over this redshift range, the $\kappa$ map is most sensitive to haloes with masses  
$\gtrsim 10^{14}M_{\odot}$ corresponding to massive galaxy clusters. These massive haloes should 
correspond to  high signal-to-noise peaks in the $\kappa$ map with a significance of 4-5$\sigma$ for our data. \citep{2004MNRAS.350..893H}.

In the actual computation of the $\kappa$ map, we use {\it gamsky2kap\_gauss\_v1.21.f}\footnote{http://th.nao.ac.jp/MEMBER/hamanatk/}, a procedure that
computes the convergence map on a regular grid based on shear data defined in the celestial (RA-Dec) frame. We use a truncated Gaussian weight \citep{2015PASJ...67...34H}
in the  \citet{1993ApJ...404..441K} inversion formula that is the basis for the map.

We evaluate the $\kappa$ field  on a regular $\theta_{\rm grid}=0.15$ arcmin grid with a truncation radius for the Gaussian of $\theta_o=15$ arcmin. We replace the integral in equation \ref{eq:kappa}
with a summation over galaxies that takes  the inverse-variance weight provided by {\it lensfit} into account. Initially we apply this procedure to the entire region shown in Figure \ref{fig:patches}.

Next we trim the region of the map to the area outlined in the dashed box in Figure \ref{fig:patches}. This region
matches the original DLS F2 footprint. This trimming does not affect the $\kappa$ map because
the truncation radius $\theta_o$ of the Gaussian is much smaller than the extent of the HSC imaging beyond the boundaries of the original F2 field.

To evaluate the significance of the convergence map,
we construct 100 individual Monte Carlo  noise maps from sets of randomly oriented lensed galaxies.
We keep the positions and absolute values of the ellipticities fixed.
If we use the Kolmogorov-Smirnov (KS) statistic to compare the random realizations, the $D$-value is 0.007$\pm$0.002. 
Below we use this error in $D$ to evaluate the significance of the signal in the
$\kappa$ map.

To assess the quality of the lensing map, we construct a $B$-mode map
by rotating each  galaxy by 45$^\circ$. Again we preserve the galaxy positions and the absolute values of the ellipticity. The resulting map is a measure of the systematics in the $\kappa$-map
because the $B$ mode signal should vanish in the weak lensing limit \citep[e.g.][]{2014ApJ...786...93U}.

To compare the $B$-mode map and $\kappa$ map with the random realizations, we again use the KS test.
The KS $D$ values for the B-mode and $\kappa$ maps are  $0.008\pm0.003$ and $0.034\pm0.004$, respectively.
The $D$-value for the B-mode map is the same as for the set of random maps.
The effective number of pixels in  both maps, $N_{\rm pix}(\sqrt{2}\theta_{\rm grid}/\theta_g)^2=26,369$. Thus
the probability of drawing the B-mode map from the random map is $\sim$36\%;
in contrast the probability of drawing the $\kappa$ map from the random map is 
$\sim 10^{-8}$\%.
Thus we can conclude that the  B-mode map is consistent with being drawn from the same distribution as the random map.
On the other hand, this test of the $\kappa$ map rejects the null hypothesis
at the $\sim6\sigma$ level.

\begin{figure}[htbp] 
   \centering
	   \includegraphics[bb=0 0 580 435, width=4in]{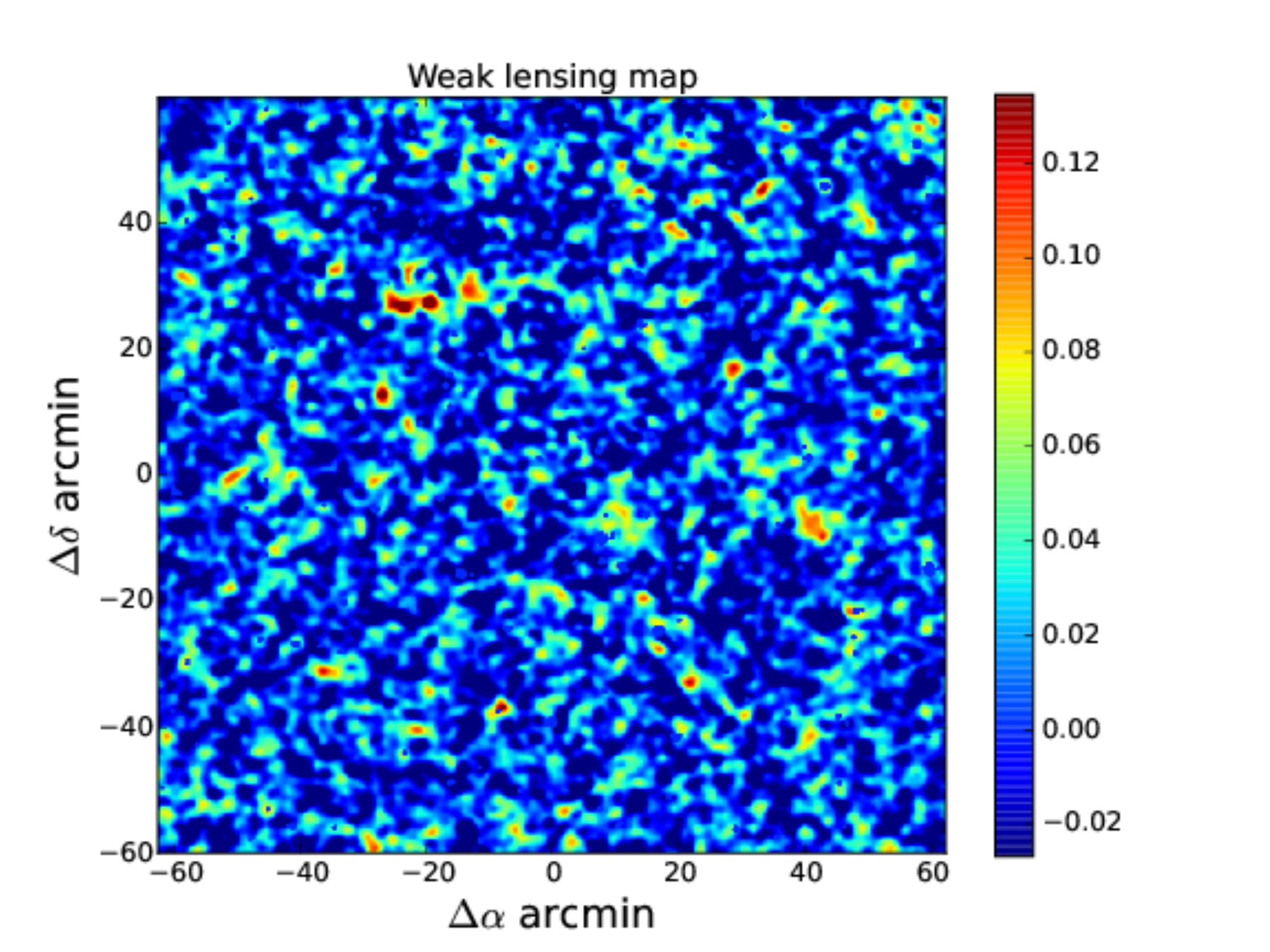}
   	   \includegraphics[bb=0 0 580 435, width=4in]{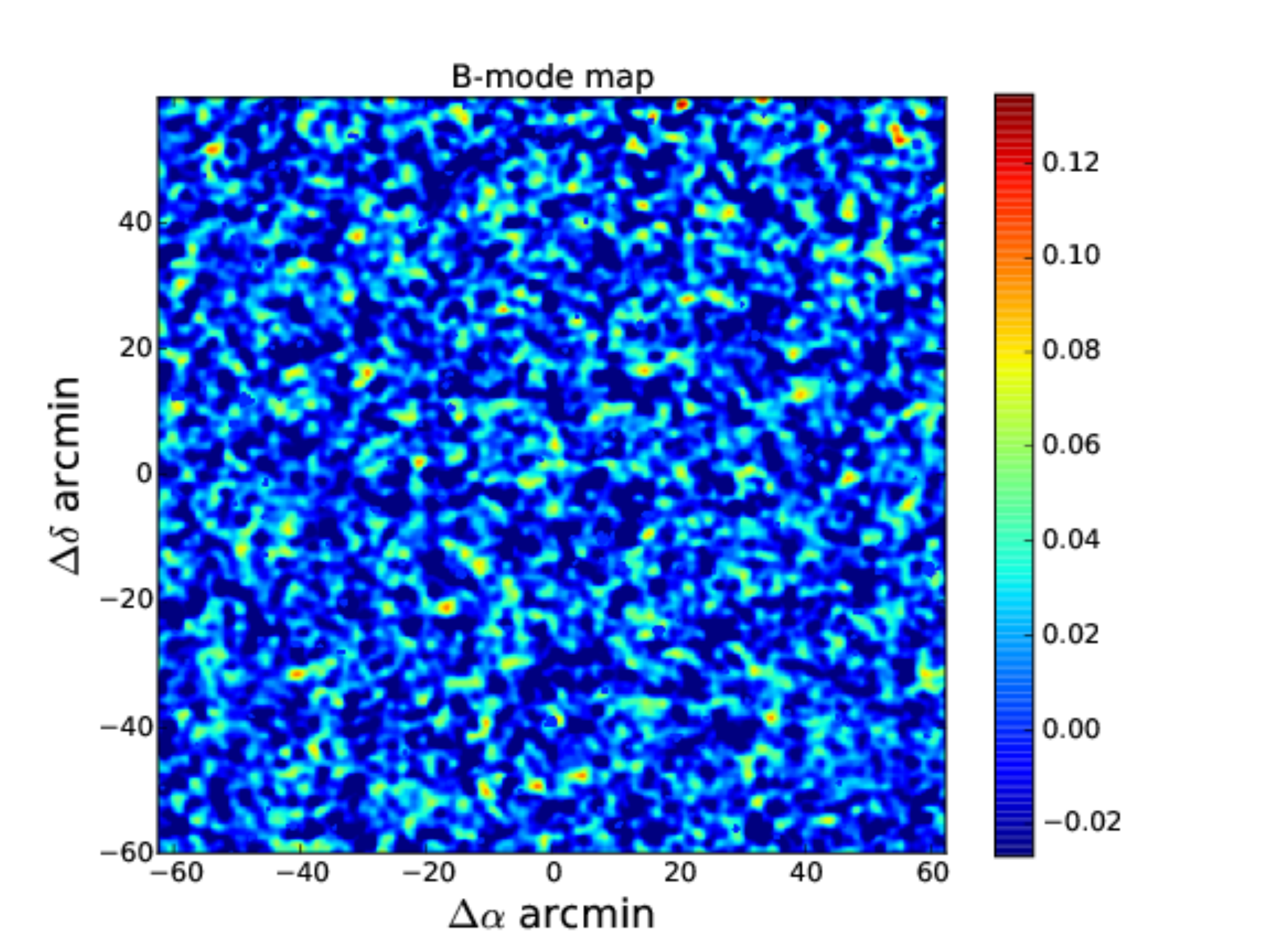}
	      \caption{The HSC weak lensing ($\kappa$) map (top) and the corresponding B-mode map (bottom). The color bar indicates the $-1\sigma$ to $5\sigma$ range for both maps. Note the obvious peaks in the lensing map and the essentially featureless B-mode map.  }
   \label{fig:wlmap}
\end{figure}

Figure \ref{fig:wlmap} shows the $\kappa$ and B-mode maps. It is visually apparent that there
are several obvious peaks in the $\kappa$ map; there are none in the B-mode map.

The KS statistic tests only the distribution of pixel values; it does not test for
correlations among the values on the sky. To test for these spatial correlations, we cross-correlate the maps.
We calculate the normalized cross-correlation function
according to the definition \cite[e.g][]{2002dip..book.....G}:
\begin{eqnarray}
	NCC(x,y) = \frac{\sum_{i,j} M_{1}(i,j) M_{2}(i+x,j+y)}{\sqrt{\sum_{i,j} [M_{1}(i,j)]^2}\sqrt{\sum_{i,j} [M_{2}(i,j)]^2}}\label{NCC}
\end{eqnarray}
where $M_1(i,j)$ and $M_2(i,j)$ are the maps. 
Once we compute the cross-correlation map, we  azimuthally average the radial profile
and denote it as $\langle C_{1;2} \rangle_r$,  where $r=\sqrt{x^2+y^2}$.
The scaling for each $M(i,j)$ does not matter because the cross-correlation is normalized by the rms value of each map. 
The cross-correlation provides a  measure
of the morphological similarity of the maps.
The zero-lag value of the cross-correlation, $\langle C_{1;2} \rangle_{r=0}$, is often used
to characterize the measure.

\begin{figure}[htbp] 
   \centering
	   \includegraphics[bb=0 0 580 435, width=4in]{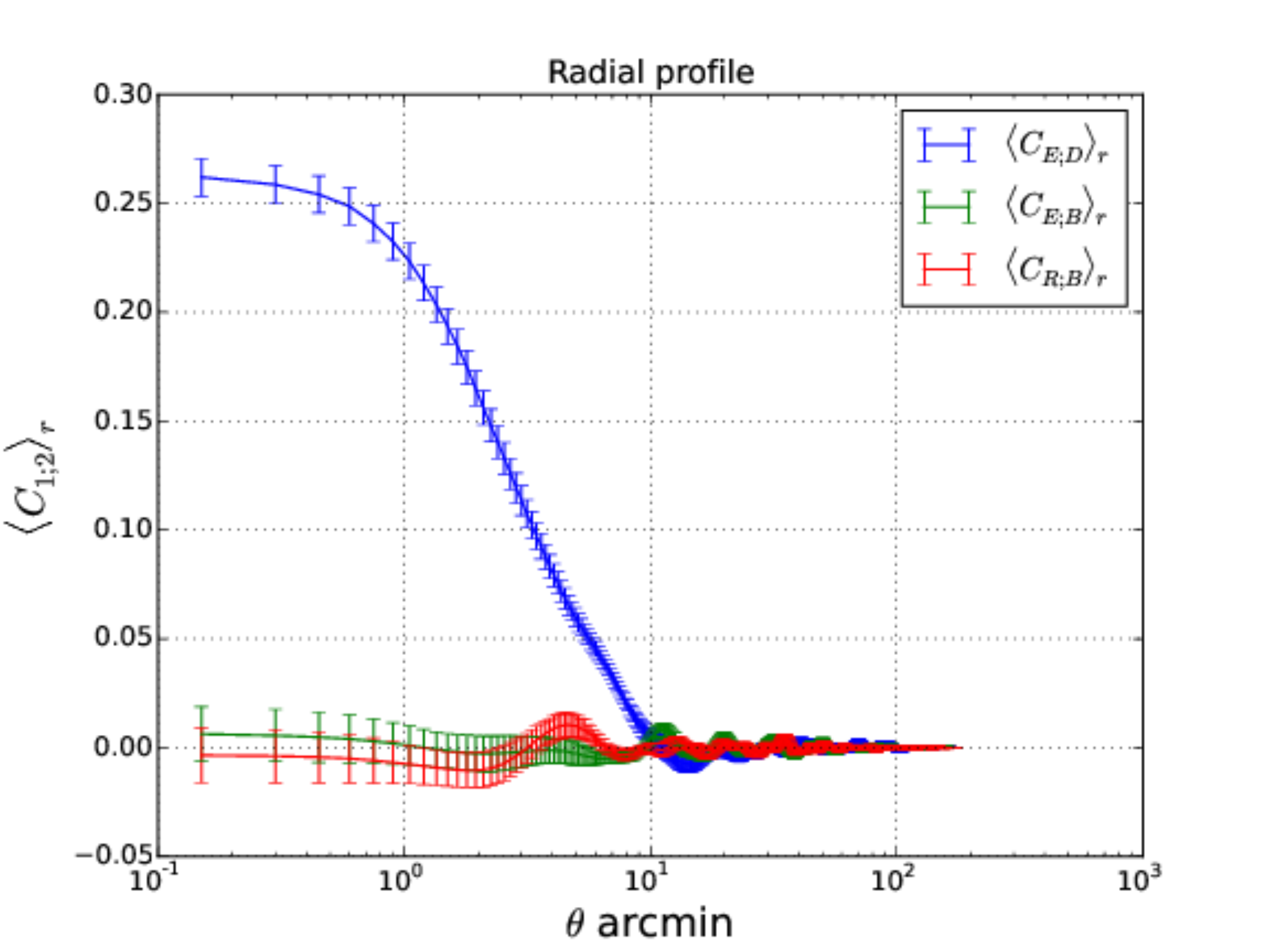}
   \caption{
 The green and red lines show the radial profile of the normalized cross-correlation between the $\kappa$ and B mode map and the kappa and noise maps, respectively. These
profiles provide measures of the noise in the normalized cross-correlation.
The blue line shows the radial profile of the normalized cross correlation between the kappa map and the predicted map (See Section \ref{morphology}). The  normalized cross-correlation between the $\kappa$ and predicted maps is significant at the $\sim 30\sigma$ level.
}
   \label{fig:1dcrosscorrelation}
\end{figure}
Figure \ref{fig:1dcrosscorrelation} shows the cross-correlation between the $\kappa$ and B-mode maps. We also show the cross-correlation between the B-mode map and a noise map.
We calculate the error bars based on 10 sets of noise maps. 
The cross-correlation amplitude  $\langle C_{E;B} \rangle_r$ is consistent with a null signal to within the error. These tests confirm that the potential systematics in the weak lensing ($\kappa$) map are negligible for our purposes.

\subsection{The predicted map}\label{predictedmap}
We next construct  a projected lensing efficiency weighted galaxy stellar mass density map in order to compare 
the matter distribution traced by the $\kappa$ map with  the matter distribution traced by galaxies in the redshift survey. The median depth of the redshift survey is $ z = 0.3$ (G14). The redshift survey traces structure to $z \sim 0.6$ and is thus reasonably well-matched to the weak lensing sensitivity. Hereafter we call this map derived from the redshift survey the predicted map.

\citet{2013MNRAS.433.3373V} developed a similar approach based on photometric redshifts
and \citet{2014ApJ...797..106H} used the same approach to compare cluster weak lensing maps with dense redshift surveys. Here we cross-correlate a predicted map derived from a dense redshift survey with the $\kappa$ map.
We use the stellar mass as a proxy for the galaxy halo mass.
Although the ratio of halo mass to stellar mass varies with stellar mass and galaxy type,
we begin by assuming a fixed overall ratio.

The spectroscopic sample is magnitude limited. Thus at greater redshift the survey encompasses sets of increasingly luminous and correspondingly massive galaxies. In order to construct a map that reflects the matter distribution traced by the galaxies, we  weight each galaxy by the un-sampled lower mass end of the mass function (see \cite{2015arXiv150803346D}). This weight is obviously a function of redshift.

We modify the weighting procedure of \cite{2015arXiv150803346D} 
to account for (1) the  scatter in the mass limit corresponding to the stellar mass limit,
and (2) we construct separate weights for quiescent and star-forming galaxies.
The mass functions from \citet[hereafter I13]{2013A&A...556A..55I} are the basis for our computation.
They fit double Schechter functions to UltraVISTA survey data to derive the mass function for $\log(M_{\star}/M_{\odot})>9$ and $z<4$.
We use the double Schechter function parameters from their Table 2
for the  $0.2<z<0.5$ interval appropriate to the bulk of our data. 

I13 derive stellar masses  using the {\it LePhare} code with a \cite{2003ApJ...586L.133C} IMF,
the same procedure we apply to the spectroscopic sample (G14).
There is a small offset between our mass estimates and I13. For 2,300 galaxies in the COSMOS field,
we calculate the stellar mass using the same parameters and models we used for the F2 stellar masses in G14.
We cross-match these 2,300 galaxies with the I13 catalog and compare the two stellar mass estimates.
The galaxies we use for the comparison have a magnitude distribution similar to the entire F2 sample.
All of the comparison galaxies have Hectospec spectroscopy,  but in the I13 analysis they use photo-zs for many of these galaxies because  these redshifts were not available.
The data are nearly normally distributed, but
we clip the $\gtrsim 3\sigma$ tail resulting from the photo$z$s used by I13.
The mean and standard error of the difference in stellar masses:
\begin{eqnarray}
\log M_{\rm Ilbert} - \log M_{\rm F2} = -0.039 \pm 0.003.	\label{eq:diff}
\end{eqnarray}
where $M_{\rm Ilbert}$ is the stellar mass from I13 and $M_{\rm F2}$ is the F2 value.
 This difference is independent of the stellar mass.
To calibrate the I13 mass function to the F2 data,  we shift the values of the characteristic masses in I13 by the small difference in equation \ref{eq:diff}.
The effect on our analysis is negligible but we make a correction for consistency.

\begin{figure}[htbp] 
   \centering
   	   \includegraphics[bb=0 0 580 435, width=4in]{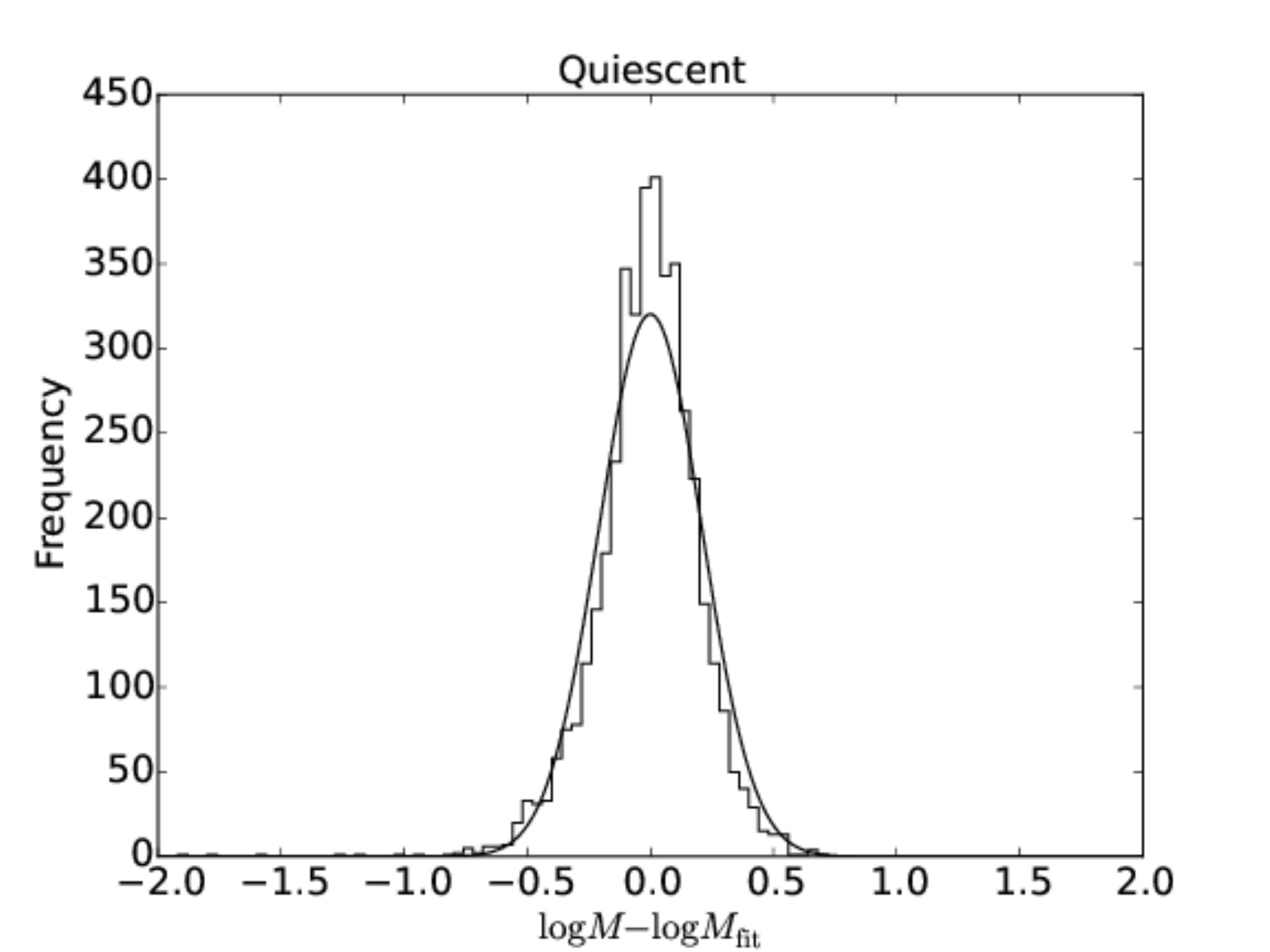}
   	   \includegraphics[bb=0 0 580 435, width=4in]{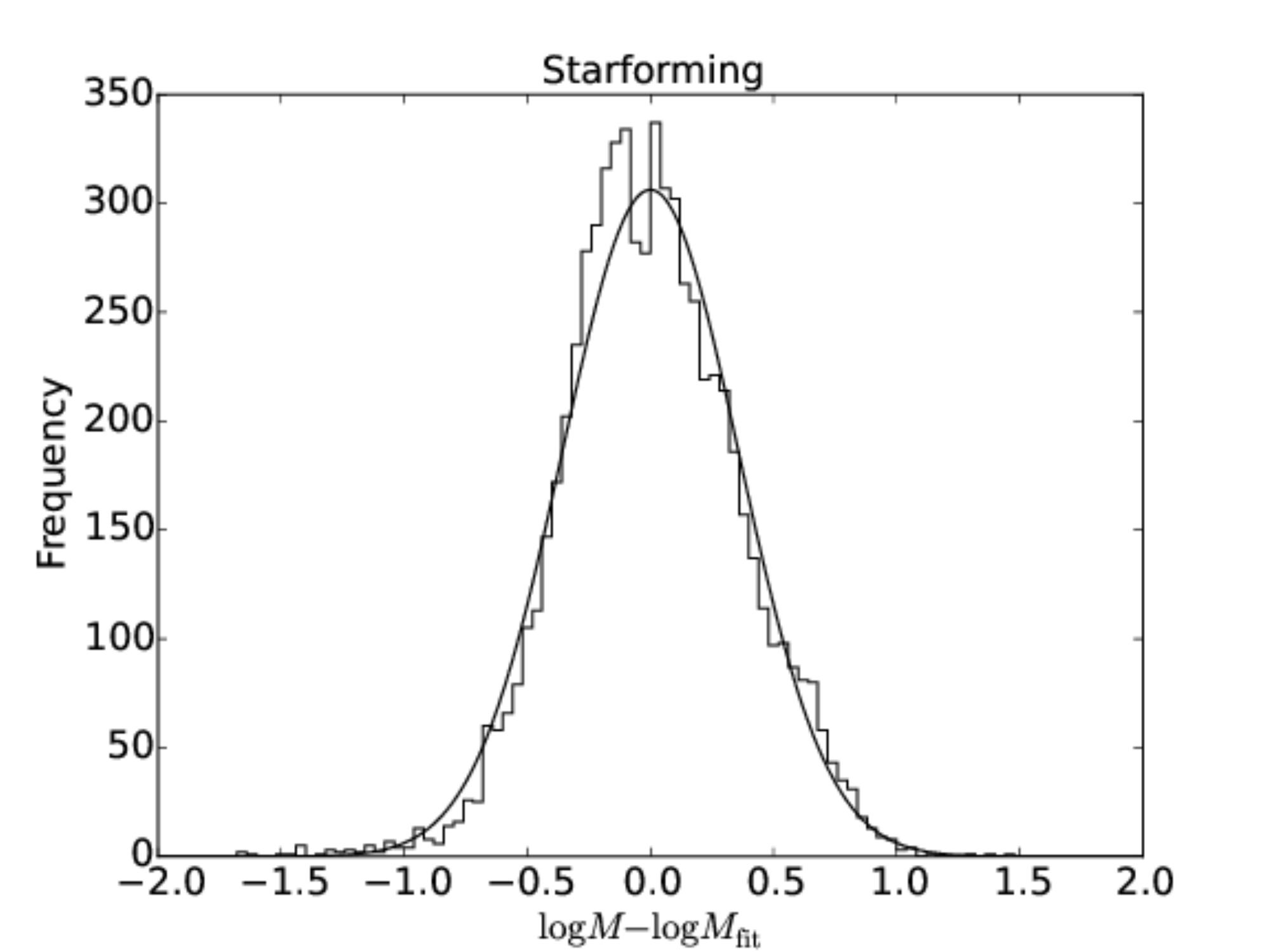}
   \caption{Distributions of residuals from fitting the log of the limiting stellar mass, $M_{\rm lim}(z)$, to the limiting $K$-corrected absolute magnitude  (Figure \ref{fig:specsamples})
for quiescent (top) and star-forming (bottom) galaxies. The curve shows  Gaussian fits to the residuals used to derive the error functions in Equation \ref{eq:wz}.} 
   \label{fig:sigma_on_mass_to_luminosity}
\end{figure}

We compute the redshift dependent weight $W_z$:
\begin{eqnarray}
W_z &=& \int_{\log M_l}^{\log M_u} \log M_{\star}\Phi(\log M_{\star}) d\log M_{\star} \nonumber\\
&/& \int_{\log M_l}^{\log M_u} \log M_{\star}\Phi(\log M_{\star})\frac{1}{2} \nonumber\\
&\times&\left[{\rm erf} \left(\frac{\log M_{\star}-\log M_{\rm lim}(z)}{\sqrt{2}\log\sigma}\right) +1 \right] d\log M_{\star}, \label{eq:wz}
\end{eqnarray}
where $\Phi$ is the Schechter function, ${\rm erf}(x) = \frac{2}{\sqrt{\pi}}\int_{0}^{x} e^{-t^{2}}dt$ is the error function and
$\log M_i=\log (M_i/M_{\odot})$ with $i=(\star,{\rm lim},u,l$) are the logarithms of the stellar masses in  solar units.
The solid red curves in Figures \ref{fig:specsamples} show the stellar mass limits,
$M_{\rm lim} (z)$ that correspond to the $K$-corrected magnitude limits for quiescent (upper panel) and star-forming (lower panel) galaxies, respectively.
The dashed lines in these figures show the scatter around the limit. The scatter $\log \sigma=\log(\sigma_{\star}/M_\odot)$ for quiescent and for star-forming is 0.21 and 0.36, respectively.
Figure \ref{fig:sigma_on_mass_to_luminosity} shows the residuals from the linear fit between the log of the stellar mass and the limiting $K$-corrected absolute magnitude for the redshift survey.
The straightforward linear relation between the log of the stellar mass and absolute magnitude
produces  residuals that follow a Gaussian distribution
supporting the adequacy of the linear fit
and the error function introduced in Eq (\ref{eq:wz}).
 
\begin{figure}[htbp] 
   \centering
	   \includegraphics[bb=0 0 580 435, width=4in]{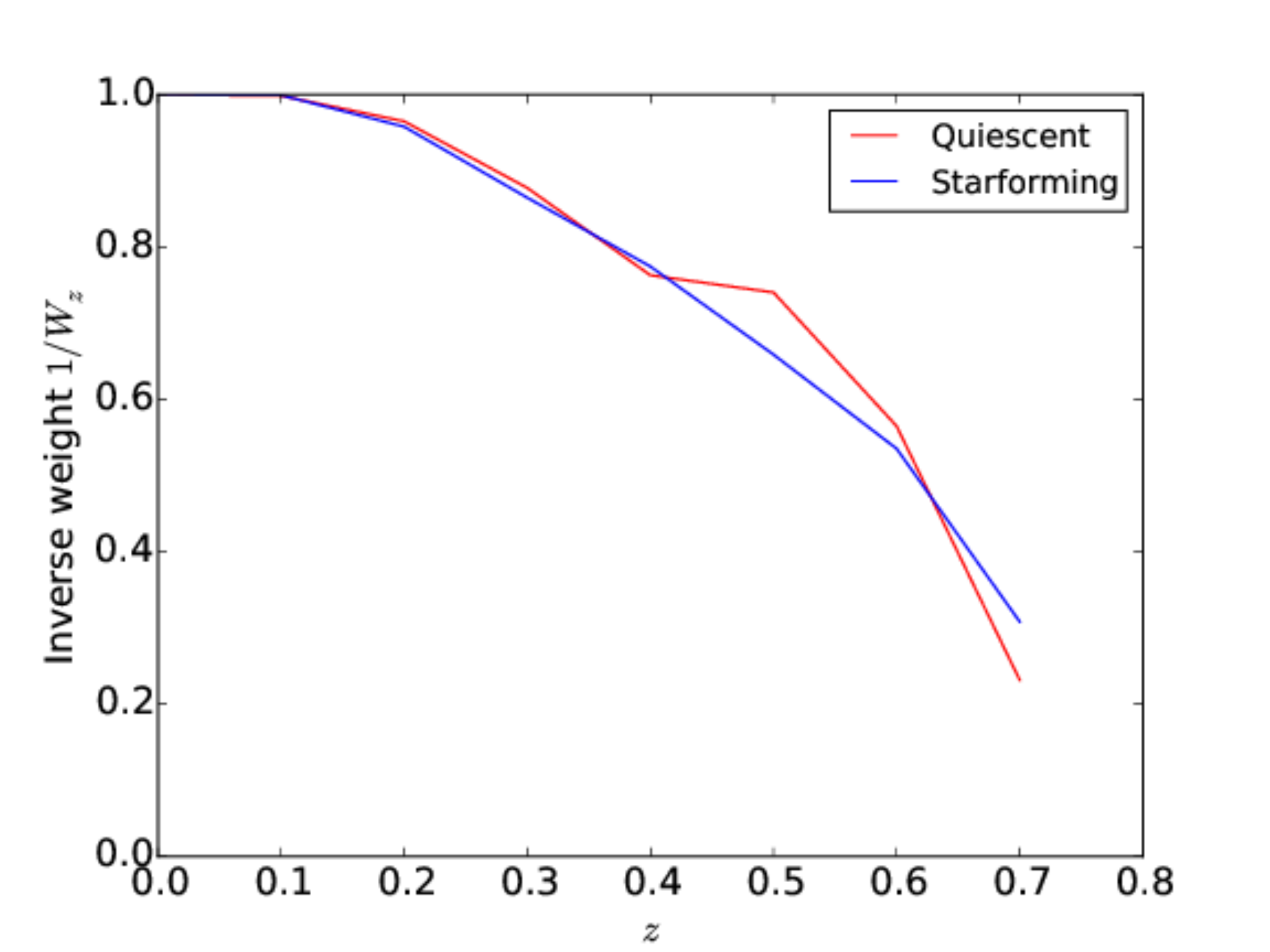}
  	   \includegraphics[bb=0 0 580 435, width=4in]{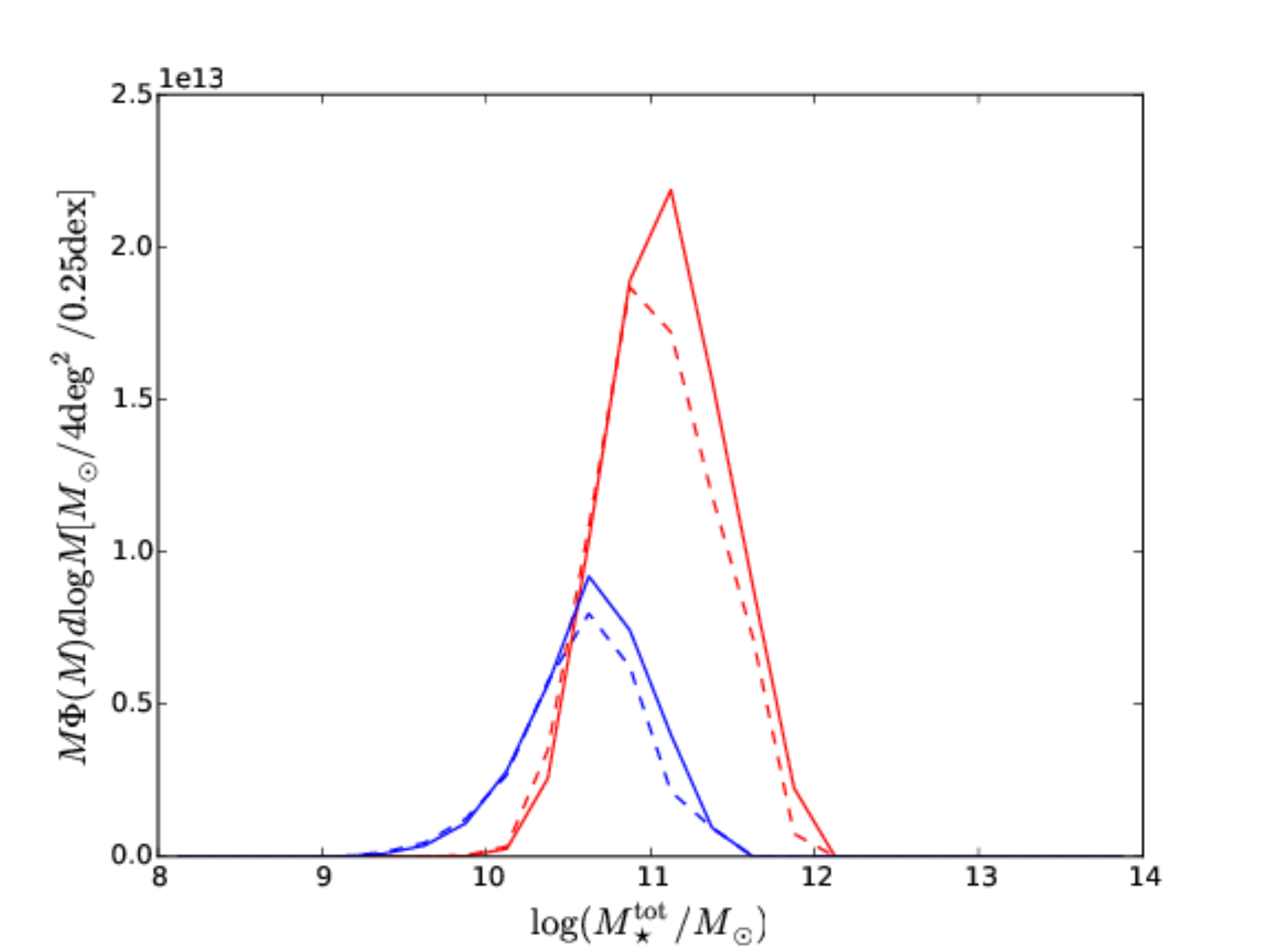}
	      \caption{Redshift dependent inverse weight $1/W_z$ (upper panel) for the quiescent (red) and 
star-forming (blue) samples from Equation \ref{eq:wz}.
Relative mass weighted contribution from each galaxy component (lower panel) for the redshift range 0.3 to 0.35, near the median depth of the redshift survey. Note that the
weighted distributions  peak in the mass range $10^{10.5}$ to 1$0^{11}M_\odot$ and that the peaks are narrow with a dispersion of $\sim$ 0.4dex.
The solid lines show the contribution weighted according to  Equation \ref{eq:wz}; the dashed lines show the directly observed contribution.} 
   \label{fig:correctionfactor}
\end{figure}

Equation \ref{eq:wz} is the ratio between the total stellar mass density in the range
$\log M_l=\log(M_{\star}^l/M_\odot)$  to   $\log M_u=\log(M_{\star}^u/M_\odot)$  that we observe (numerator) to the observed portion of the mass function taking the error in the limiting stellar mass into account (denominator).
We adopt $\log (M_{\star}^l/M_{\odot})$ = 8 and $\log (M_{\star}^u/M_{\odot})$ = 13 to cover the observed range (Figure \ref{fig:specsamples}).
Figure \ref{fig:correctionfactor} shows the  inverse of the weight $1/W_z$ as a function of redshift. The estimated effective total stellar mass for a galaxy of mass $M_*$ in the magnitude-limited survey is 
\begin{eqnarray}
M_{\star}^{\rm tot}= M_{\star}\times W_z
\end{eqnarray}
with the appropriate $W_z$ for quiescent and star-forming galaxies.

The predicted map is a sum of the shear signals
produced by each individual lensing galaxy in the foreground redshift survey
and covers the effective redshift range $0.1\lesssim z\lesssim0.7$
(Eq.  \ref{eq:sumofshear}):
\begin{eqnarray}
	\kappa_{\rm gal} (\boldsymbol{\theta})= \sum_{i=1}^N \frac{\Sigma_i(|\boldsymbol{\theta}-\boldsymbol{\theta_i}|)}{\Sigma_{\rm crit}(z_{l}^{(i)})}-\bar{\kappa}_{\rm gal} \label{eq:sumofshear}
\end{eqnarray}
where $\Sigma_i$ is the shear signal from the $i$-th individual lensing galaxy,
$\Sigma_{\rm crit}(z_{l}^{(i)})$ is the critical surface mass density for the known lens redshift $z_l^{(i)}$
for the $i$-th galaxy, and $\bar{\kappa}_{\rm gal}$ is the mean mass density.

The shear signal from each galaxy in the redshift survey can be calculated if the position on the sky, the distance, and the mass profile are known.
We assume an NFW mass profile  characterized by the total mass and concentration \citep{2000ApJ...534...34W}. We tacitly assume that the total mass of each galaxy is 
proportional to the stellar mass. 
We assign an NFW shear profile to each galaxy in the  spectroscopic sample with $z \leq0.7$.
At greater redshift the redshift survey is very sparse and the weight function is large.



We adopt the mass-concentration relation calibrated with numerical simulations in \citet{2011MNRAS.411..584M}.
To obtain the concentration parameter, we simply assume a constant stellar-halo mass ratio
$M_{h}/M_{\star}\sim50$ for both populations. We base this approach on the weak lensing
results of \citet{2014MNRAS.437.2111V}. They obtain
$M_{h}/M_{\star}\sim60$ for quiescents with $10^{11}M_{\odot}$ and $\sim40$  for star-forming with $10^{11}M_{\odot}$. The weak lensing assessment of the stellar mass to halo mass ratio by \citet{2014MNRAS.437.2111V} applies for the redshift range 0.2-0.4. Figure 8 (lower panel) shows that these values apply to the galaxies
that dominate the predicted map.
At this stage of the construction of the predicted map we ignore any
differences between the halo mass to stellar mass ratios for quiescent and star-forming galaxies.
We discuss this issue further in Section \ref{limitation}.

To calculate the convergence for each lensing galaxy,
we compute the critical surface mass density 
\begin{eqnarray}
	\Sigma_{\rm crit} &=&\frac{c^2 }{4\pi G D_l} \left\langle\frac{D_{ls}}{D_s} \right\rangle^{-1}
\end{eqnarray}
where 
\begin{eqnarray}
	\left\langle\frac{D_{ls}}{D_s} \right\rangle = \int_{zl}^{\infty} dz n_z \frac{D_{ls}}{D_s}/\int_{zl}^{\infty} dz n_s(z),
\end{eqnarray}
and where $D_l$, $D_{ls}$, and $D_s$ are angular diameter distances to the lens, to the source and between the lens to the source.
We parameterize the redshift distribution of the sources according to
\begin{eqnarray}
	n_s(z) = \frac{\beta}{z_{\star}\Gamma[(1+\alpha)/\beta]}\left(\frac{z}{z_{\star}}\right)^{\alpha}
	\exp\left[-\left(\frac{z}{z_{\star}}\right)^{\beta} \right]
\end{eqnarray}
\citep{2009PASJ...61..833H}.
The mean redshift of the source galaxies is  $\langle z_s\rangle = z_{\star}\Gamma[(2+\alpha)/\beta]/\Gamma[(1+\alpha)/\beta]$.
We assume $\alpha=1.5$ and $\beta=1$ following \citet{2009PASJ...61..833H} and we adjust 
the mean redshift to our dataset, $\langle z_s \rangle=1.12$ (see Section \ref{shearcatalog}).

As discussed in Section \ref{SHELSDESCRIPTION}, we must mask a small region of the survey to remove regions affected by bright stars. The mask covers $\lesssim 5$\% of the survey area and is the same for the $\kappa$ map and for the predicted map. Here we replace the masked regions with  zero values. Finally we smooth the map with a kernel width of 1 arcmin, identical to the $\kappa$
map. 

\begin{figure}[htbp] 
   \centering
	   \includegraphics[bb=0 0 580 435, width=4in]{fig5a.pdf}\\
   	   \includegraphics[bb=0 0 580 435, width=4in]{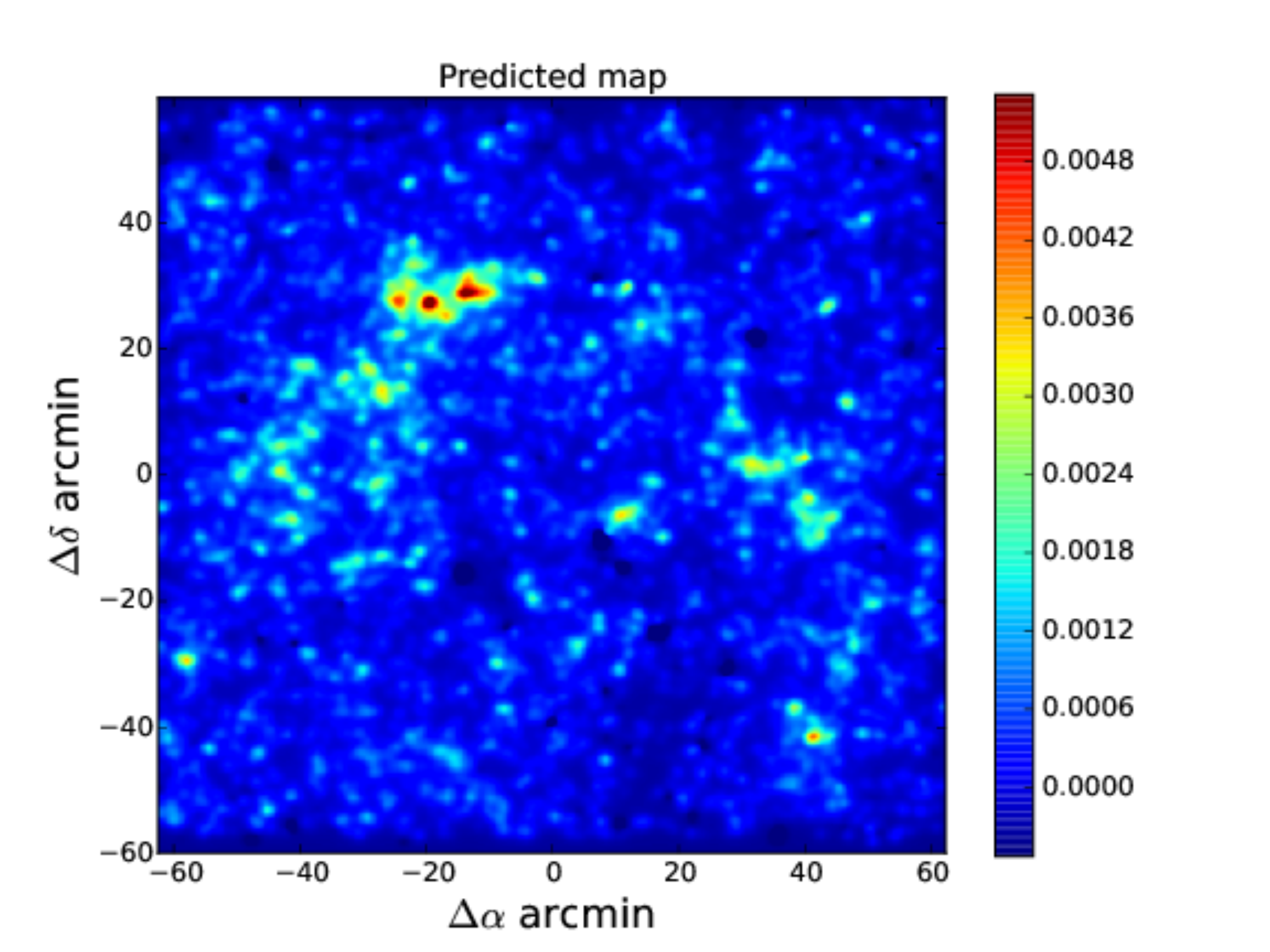}
	      \caption{The HSC $\kappa$ map (top) and the predicted map (bottom). The color bar for the $\kappa$ map covers the range $-1\sigma$ to $5\sigma$. For the predicted map the color bar shows the range -1$\sigma$ to 10$\sigma$. Note, for example, the correspondence of peaks at $(\Delta\alpha, \Delta\delta)\sim(-20^{\prime}, -30^{\prime})$ and the essentially signal-free region centered at $(\Delta\alpha, \Delta\delta)\sim(-10^{\prime}, -20^{\prime})$ and $(30^{\prime}, -30^{\prime})$.}
   \label{fig:predictedmap}
\end{figure}

Figure \ref{fig:predictedmap} shows the resulting map. Visual comparison of the predicted map with the $\kappa$ map reveals the morphological similarity.
For example,  the concentration of  3 peaks at $(-20^{\prime}, 30^{\prime})$ 
and a chain of structures from ($40^{\prime}, 0^{\prime}$) to ($40^{\prime}, -50^{\prime}$) are nearly identical. 
A large region centered at $(-10^{\prime}, -20^{\prime})$ and $(30^{\prime}, -30^{\prime})$ has essentially no structure in either map.

\section{Cross-Correlating the Maps}\label{sec:crosscorrelation}

Cross-correlation of the $\kappa$ and predicted maps provides a quantitative measure of the correspondence between the structures revealed by the two maps (Figure \ref{fig:predictedmap}).
By construction, both maps are smoothed on the same scale, 1 arcminute.  
Shape noise is intrinsic to the $\kappa$ map, but we do not include it in the predicted map
because it requires arbitrary assumptions (Section \ref{predictedmap}).

When we derive the predicted map, we take the lensing efficiency as a function of redshift into account directly.
The $\kappa$ map is sensitive to structure in the range 0.1$\lesssim z \lesssim 0.7$. 
However, the predicted map may not include all of the relevant structures that contribute to the lensing signal, particularly at redshifts $\gtrsim 0.6$. Although the redshift survey extends to $z \sim 0.7$,
the sampling density is so low that even massive clusters may not be sampled well enough to appear as significant peaks in the predicted map.

In Section \ref{morphology} below we investigate the contribution of individual redshift slices to the cross-correlation signal.
Finally, in each redshift slice, Section \ref{starforming} examines the matter distribution  
traced by massive quiescent galaxies  and by massive star-forming galaxies.

        \subsection{Comparing the $\kappa$ Map and the Predicted Map}\label{morphology}

Figure \ref{fig:1dcrosscorrelation} shows the azimuthally averaged cross-correlation (Equation \ref{NCC}) between the
$\kappa$ and predicted maps. The amplitude at zero-lag reflects the impact of shape noise on the weak lensing map as well as the failure to sample structure at high redshift. We determine the error in the cross-correlation by cross-correlating the predicted map with 
10 sets of 100 random maps and then computing the standard deviation among the radial profiles. The cross-correlation between the $\kappa$ map and the predicted map is significant at the 30$\sigma$ level.

The significance of the cross-correlation we measure is a substantial improvement over previous results \citep{1998astro.ph..9268K,2001ApJ...556..601W,2005ApJ...635L.125G,2013MNRAS.433.3373V,2015PhRvD..92b2006V}.
The main  reasons for the improvement are (1) the depth and seeing of the HSC observations and (2) the depth and density of the redshift survey.
\cite{2005ApJ...635L.125G} is the only  previous comparison based on a dense redshift survey.
They used local velocity dispersion rather than galaxy stellar mass density as a proxy for the surface mass density. They detected the cross-correlation between the $\kappa$ and velocity dispersion map at the 6.7$\sigma$ level. The redshift survey was shallower and the spatial resolution of the velocity dispersion
map was only 2.5 arcmin, comparable with the DLS map based on a source density of $\sim$ 14 arcmin$^{-2}$.

\cite{1998astro.ph..9268K} based their cross-correlation on a supercluster region using photometric properties of red galaxies only; their source number density and median seeing are similar to ours.
The significance of their cross-correlation between the $\kappa$ and photometric comparison map is $9\sigma$  for a map with a smoothing scales of  $45$ arcsec.
\cite{2001ApJ...556..601W}  cross-correlate a weak lensing map with the foreground distribution of red galaxies in a more general region, more comparable with the DLS F2 region. They detected the zero-lag cross-correlation at the $5.2\sigma$ level.
\citet{2013MNRAS.433.3373V} cross-correlated their $\kappa$ map of a $154\ {\rm deg}^{2}$ region
surveyed for the Canada-France-Hawaii Telescope Lensing Survey (CFHTLenS) with a predicted map based on photometric redshifts.
They computed the stellar mass of central galaxies and assigned a standard halo mass ratio. Their zero-lag cross-correlation has an amplitude of $\sim0.12$ for a 1.8 arcmin smoothing corresponding to a $\sim10\sigma$ significance. Most recently 
\citet{2015PhRvD..92b2006V} cross-correlated a $\kappa$ map based on $139\ {\rm deg}^{2}$ imaging of  Dark Energy Survey (DES) science verification data with the foreground distribution of galaxies  \citep{2015PhRvL.115e1301C}.
They use number density rather than mass density to characterize their foreground map. 
They also use photometric redshifts. The statistical significance of their cross-correlation is 6.8$\sigma$ level with 20arcmin smoothing.

The region we cover is only 4 deg$^2$. Nonetheless the comparison between the $\kappa$ map and the predicted map yields a 30$\sigma$ amplitude for the zero-lag cross-correlation signal.
The high signal-to-noise for this small region is a consequence of the HSC imaging and the dense foreground redshift survey.
The predicted map is undiluted by background sources because we construct it from a redshift survey rather than from a purely photometric catalog.
The average density of the redshift survey is 1 galaxy arcmin$^{-2}$.
The $\kappa$ map is probably not strongly contaminated by foreground sources because the depth of the HSC imaging enables use of exclusively faint galaxies for constructing the map.
These source galaxies are so faint that they are very unlikely to be members of foreground clusters. \cite{2010PASJ...62..811O} estimate that the foreground contamination should be $\sim 10\%$ by mass even in the background of a rich cluster.

The width of the cross-correlation is a measure of the typical scale of the large-scale structure reflected in the maps.
The full-width half-maximum is $\sim$6 arcmin, corresponding to a physical scale of $\sim$1.6 Mpc at the median depth of the redshift survey, $z \sim 0.3$.
This scale is comparable with the scale of rich clusters of galaxies and suggests that they dominate the cross-correlation as expected. 
We next explore the impact of these systems in more detail by cross-correlating the $\kappa$ map with slices through the redshift survey.

\subsection {The $\kappa$ Map and Slices of the Redshift Survey} \label{sec:zslices}
\begin{figure}[htbp] 
   \centering
	   \includegraphics[bb=0 0 580 435, width=4in]{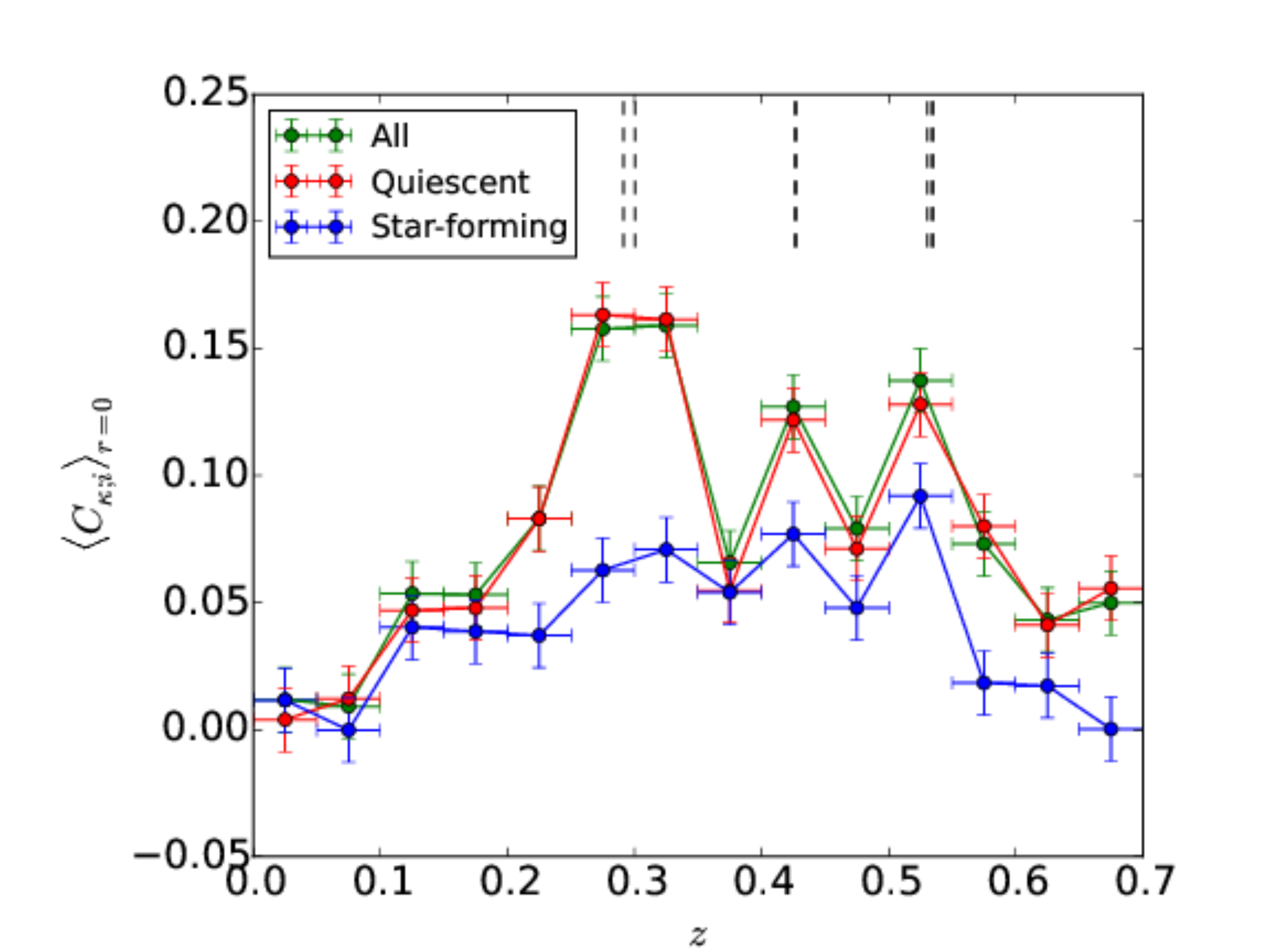}
   \caption{Zero-lag cross-correlation as a function of redshift between the $\kappa$ map and the total predicted map (green), between the $\kappa$ map and the quiescent predicted map (red) and between the $\kappa$ map and the predicted star-forming map (blue). The redshift slices have a width $dz = 0.05$. Dotted lines indicate massive independently identified X-ray clusters from \cite{ 2014ApJ...786..125S} with masses $M_{500} \gtrsim 10^{14}$M$_\odot$. The correspondence of the peaks
reinforces the importance of the contribution of these massive systems to the weak lensing signal.}
   \label{fig:corr_z.png}
\end{figure}
To explore the impact of structures at different redshifts,
we bin the redshift survey in slices with a width $dz=0.05$.
For each redshift slice, we repeat the procedure in Section \ref{predictedmap} to construct a set of  predicted maps.
We then compute the value of the zero-lag cross-correlation for each subsample.

Figure \ref{fig:corr_z.png} shows the zero-lag cross-correlation as a function of redshift.
We note that this comparison is insensitive to the redshift dependent weight $W_z$ because we use the normalized cross-correlation function defined in Equation \ref{NCC} which is essentially independent of the amplitude of the maps. 

Figure  \ref{fig:corr_z.png} shows peaks in the ranges $0.25<z<0.3, 0.3<z<0.35, 0.4<z<0.45$ and $0.5<z<0.55$.
At these redshifts, \cite{2009ApJ...702..980K}, \cite{2009ApJ...702..603A}, \cite{2010ApJ...709..832G}, \cite{2014ApJ...786..125S}, \cite{2014ApJ...786...93U}, and \cite{2015ApJ...807...22M} identify clusters in the F2 region.
Massive X-ray clusters are concentrated  at three redshifts ($z\sim0.30$ (2 clusters)  , $z \sim 0.43$ (2 clusters), and $z\sim 0.53$ (3 clusters)) and their masses  derived from X-ray observations are $M_{500} \gtrsim 1\times 10^{14}$M$_\odot$
\citep{2014ApJ...786..125S}.
All of these X-ray clusters are detected in the redshift survey and in other weak lensing  observations. 
The vertical dashed lines in Figure  \ref{fig:corr_z.png} indicate the redshifts of these massive X-ray clusters; the correspondence with the lensing peaks is impressive.
As suggested by the width of the overall cross-correlation (Figure \ref{fig:corr_z.png}), the amplitude of the zero cross-correlation is maximum in redshift intervals that contain massive clusters of galaxies. 

\begin{figure}[tbp] 
   \centering
   	   \includegraphics[bb=0 0 580 435, width=4in]{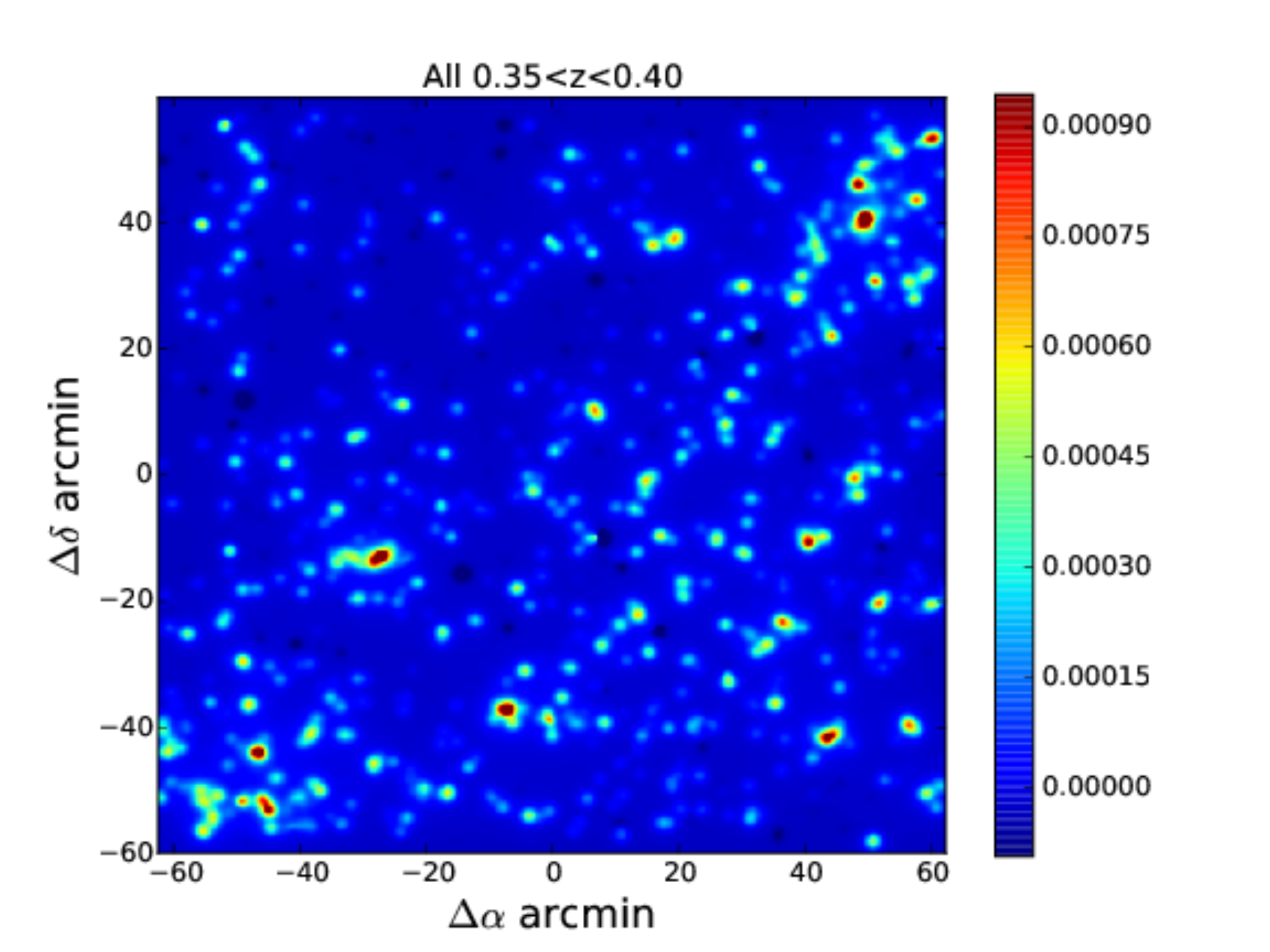}
   \caption{Redshift slice of the  predicted map  in the range $0.35<z<0.40$
where there is no peak in Figure \ref{fig:corr_z.png}.
The cross-correlation signal in this redshift range  originates for the many low lying peaks that trace the large-scale structure in this slice. Note that the range of the color bar is only $\sim 20\%$ of the range in the predicted map of Figure \ref{fig:predictedmap}.  This range corresponds to $-1\sigma$ to $10\sigma$ for this map. }
   \label{fig:nopeakslice}
\end{figure}

The zero-lag cross-correlation signal remains significant over the redshift range $0.1 <z < 0.7$ at $\gtrsim 3\sigma$ for both  the total and quiescent maps. However for the maps derived from the star-forming population at  $z\gtrsim0.6$, the zero-lag cross-correlation signal is not significant because the redshift survey  becomes  sparse.
The cross-correlation signal off the obvious peaks originates from lower mass groups that trace the large-scale of the universe.
For example, Figure \ref{fig:nopeakslice} shows  the predicted map for the redshift slice $0.35<z<0.4$. 
The many low-lying peaks correspond to groups of galaxies in the redshift range.  The situation for the redshift range $0.1 < z < 0.2$ is similar; the survey contains no massive clusters in this redshift range and is dominated by lower mass groups and large low density regions.
  
\subsection {The Contribution of Star-Forming Galaxies to the $\kappa$ Map}\label{starforming}

In previous work, most investigators have concentrated on the way quiescent red galaxies trace the matter distribution in the universe. Because we have a complete redshift survey, we can use it in combination with the $\kappa$ map to examine the contribution of massive star-forming galaxies to the weak lensing signal. This investigation is interesting because
it is now well-known that the typical  masses of star-forming galaxies decline with redshift (e.g. Figure 13 in G14)
This downsizing \citep{1996AJ....112..839C,2006ApJ...651..120B} is a more general manifestation of the original Butcher-Oemler effect \citep{1984ApJ...285..426B}
where the abundance of star-forming galaxies increases with redshift  within
the central regions of clusters at redshifts $\lesssim 0.5$.
Here we investigate the possible impact of these effects on the relationship between the $\kappa$ map and the predicted map.

We divide the galaxy sample into quiescent ($D_n4000 \geq$ 1.5) and star-forming ($D_n4000 < 1.5$) subsets.
For each subset we construct  predicted maps as described in Section \ref{predictedmap}.
We then compute the value of the zero-lag cross-correlation for each subset.

Figure \ref{fig:corr_z.png} also shows the value of the zero-lag cross-correlation
between  the $\kappa$ and the predicted maps for quiescent and star-forming subsamples as a function of redshift.
The zero-lag cross-correlation  for the quiescent subset is nearly the same as the amplitude for the total sample.
Thus quiescent galaxies are reasonable tracers of the mass distribution in the universe for $z \lesssim0.7$.
However, star-forming galaxies play an interesting role.
In general Figure \ref{fig:corr_z.png} shows that the zero-lag cross-correlation amplitude between the $\kappa$ 
and the star-forming predicted maps increases  with redshift in bins where there are massive clusters.
The significance of the cross-correlation for the star forming galaxies increases from $\sim5\sigma$ at $0.25<z<0.3$ to $\sim7\sigma$ at $0.5<z<0.55$.

\begin{figure}[tbp] 
   \centering
   \gridline{
   	   \includegraphics[bb=0 0 580 435, width=4in]{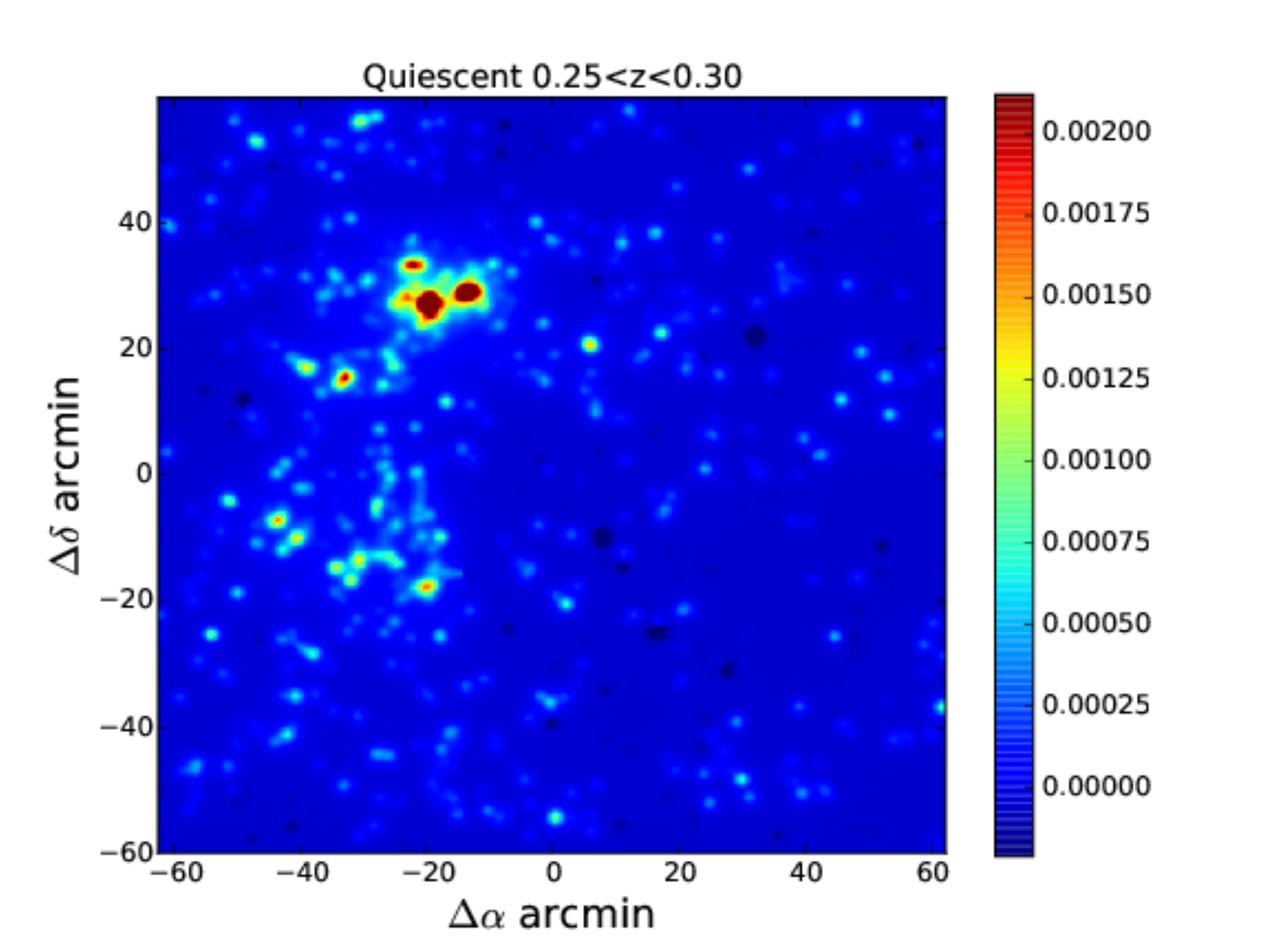}
	   }
   \gridline{
	   \includegraphics[bb=0 0 580 435, width=4in]{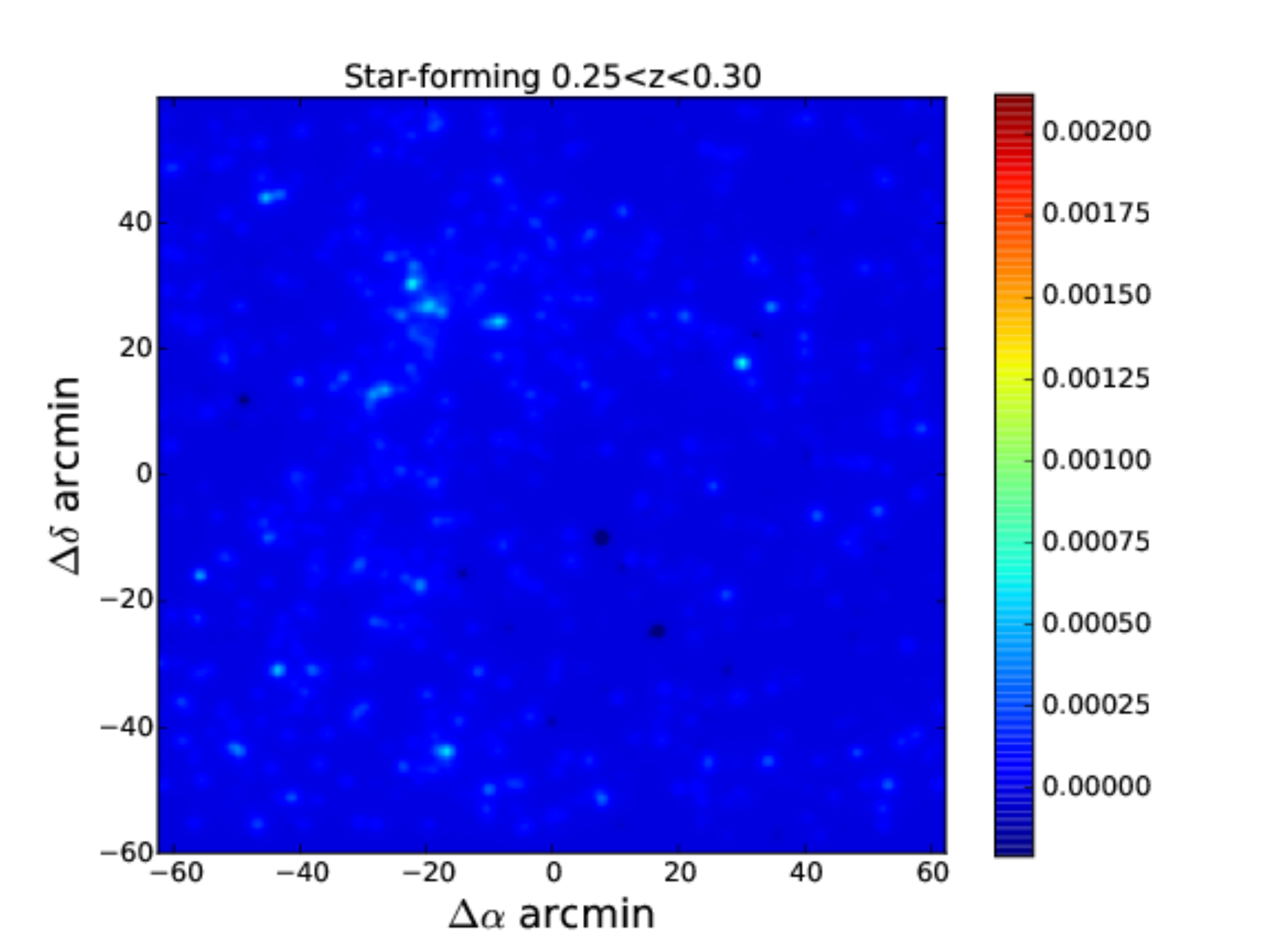}
	   }
   \caption{Redshift slices of the predicted map for quiescent (top) and star-forming (bottom)  galaxies in the redshift range $0.25<z<0.30$  containing the rich cluster complex A781 $(\Delta\alpha, \Delta\delta)\sim(-20^{\prime}, -30^{\prime})$. Note the prominence of the cluster in the quiescent map and its near absence in the star-forming map. This behavior corresponds to the ratio of cross-correlation amplitudes in Figure \ref{fig:corr_z.png}. }
   \label{fig:propertiesForLowRedshift}
\end{figure}
\begin{figure}[tbp] 
   \centering
   \gridline{
   	   \includegraphics[bb=0 0 580 435, width=4in]{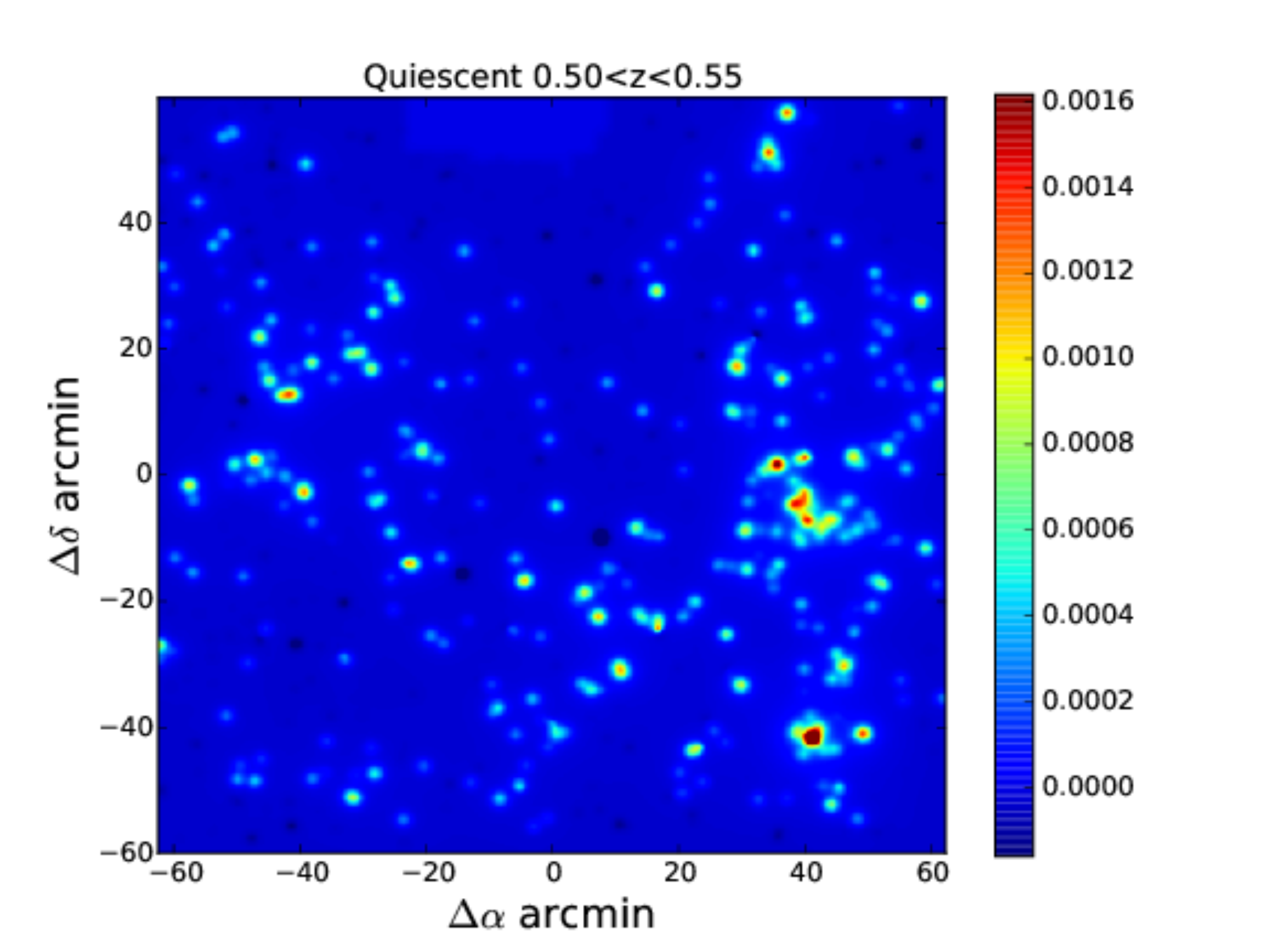}
	   }
   \gridline{
	   \includegraphics[bb=0 0 580 435, width=4in]{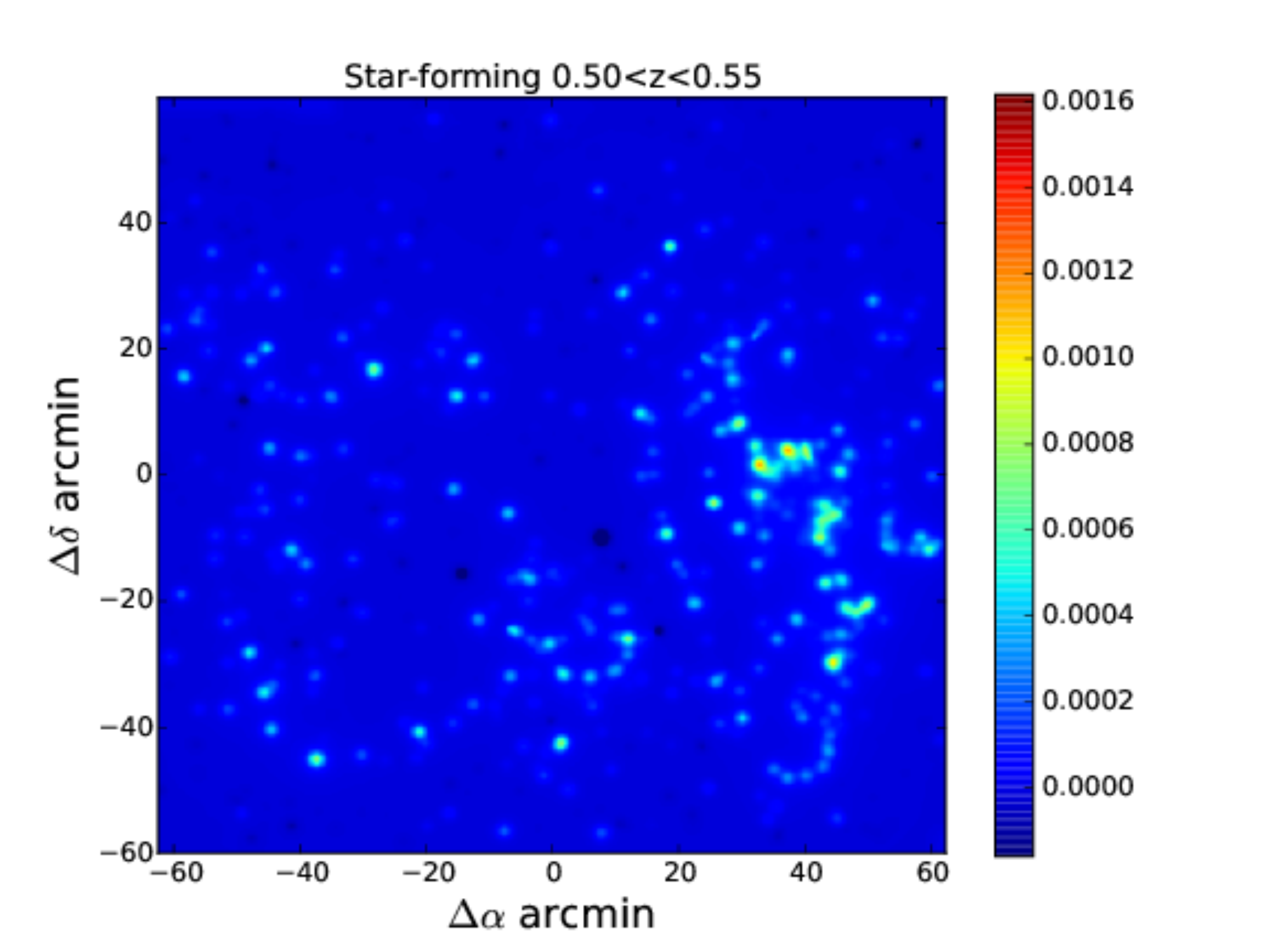}
	   }
   \caption{Same as Figure \ref{fig:propertiesForLowRedshift} but for the $0.50<z<0.55$
slice that also contains rich clusters. The upper map is for quiescent galaxies and the lower map is for star-forming galaxies. Here the clusters and their surroundings are evident in both maps. The corresponding amplitude of the cross-correlation for the star-forming map is correspondingly a larger fraction of the total signal in Figure \ref{fig:corr_z.png}.}
   \label{fig:propertiesForHighRedshift}
\end{figure}
Two examples of predicted maps at $0.25<z<0.3$ and $0.5<z<0.55$ for quiescent and star-forming (Figure \ref{fig:propertiesForLowRedshift} and \ref{fig:propertiesForHighRedshift})
underscore the striking relative behavior of quiescent and star forming galaxies as tracers of the matter distribution  in slices where there are massive systems.

In Figure \ref{fig:propertiesForLowRedshift}, there are two prominent peaks at ($-20^{\prime}, -30^{\prime}$) in the quiescent map corresponding to the components of the rich cluster A781 at $ z \sim 0.30$. These peaks are hardly visible in the star forming map.

In Figure \ref{fig:propertiesForHighRedshift}, a chain of structures from ($40^{\prime}, 0^{\prime}$) to ($40^{\prime}, -50^{\prime}$) is visible clearly in both the quiescent and star-forming maps with somewhat different morphology. There are three massive X-ray clusters in this region \citep{2014ApJ...786...93U,  2014ApJ...786..125S}. Star-forming galaxies  populate both the clusters and the surrounding region.

The result here is related to the increasing fraction of massive star-forming galaxies as a function of redshift (Figure \ref{fig:fraction}),
but it is an independent measurement of their impact as tracers of the large-scale matter distribution in the universe.
The normalized cross-correlation measures the morphological similarity of the distributions of massive quiescent and star-forming galaxies.
The correspondence between the distributions increases with redshift  in regions surrounding massive clusters. In regions dominated by groups of galaxies the signal in the lensing map is similar for star-forming and quiescent galaxies throughout the redshift range.
Although red galaxies are good tracers of the matter distribution for $z \lesssim$ 0.7,
these results imply that massive star-forming galaxies must be taken into account for larger redshifts.
The comparison among these maps provides a first weak lensing view of the downsizing of star-forming galaxies and their relationship with the large-scale structure.

\section {Discussion}\label{sec:discussion}

\subsection {Limitations of the $\kappa$ Map Comparison}\label{limitation}

Cross-correlation between the $\kappa$ and predicted maps is a powerful tool for exploring 
the relationship between the galaxy and matter distributions as a function of redshift and galaxy properties.
However the comparison is limited to relative rather than absolute measures because we lack an absolute calibration for the maps. 

We construct the predicted map by assuming that the redshift distribution of the lensed sources matches a parametric model based on photometric redshifts from VIRMOS-DESCARTES \citep{2002A&A...393..369V,2006A&A...452...51S}.
In this model the mean source redshift is 1.12.
(Sections \ref{shearcatalog}, \ref{predictedmap}).
Ambiguity in the mean source redshift and in the distribution around the mean act as a constant multiplicative bias.
As a result of this limitation extraction of the galaxy stellar-to-halo mass ratio from the relationship between the $\kappa$ map and the predicted map is not currently possible with these data.

In constructing the predicted map we use stellar mass as a proxy for the total galaxy mass
because we can measure the stellar mass from the combination of SDSS photometry and the
F2 redshift survey data.
When we cross-correlate the maps we effectively assume a constant stellar-to-halo mass  ratio for all of the objects that contribute to the map. 

We represent each galaxy with an NFW profile with a concentration appropriate to the typical stellar-to-halo mass ratio for objects in the stellar mass range covered by the redshift survey.
The predicted map is insensitive to variations in this choice because the smoothing scale of the map is of 1 arcmin, vastly larger than the scale of an individual galaxy.

To segregate the galaxy populations we use the spectral indicator $D_n4000$.
This indicator has been used by many previous investigators to demonstrate the downsizing of the star-forming galaxy populations \citep{2006ApJ...651..120B,2006MNRAS.373..349R,2010A&A...524A..67M,2013A&A...558A..61M}.
\citet{2013A&A...558A..61M} investigate the sensitivity of the sample of star-forming galaxies to definitions based on spectroscopic or photometric indicators.
They show that the segregation of massive galaxies like those we consider is insensitive to the particular method of identifying the star-forming population.

Because the redshift survey is magnitude limited, it becomes progressively more sparse at greater redshift.
In effect the spatial resolution of the predicted map is low for redshifts $\gtrsim 0.6$.
This limitation means that there may well be structure in the weak lensing map that has no counterpart in the predicted map simply because the predicted map does not cover a sufficiently large redshift range.
This limitation is actually  apparent in Figure \ref{fig:corr_z.png} where the signal becomes insignificant at $z\sim 0.6$.
This behavior reflects the limitation of the predicted map and is unlikely to be astrophysical.
A deeper redshift survey densely sampling the range $0.6 \lesssim z \lesssim 1$ could provide a more complete view of the contribution of star-forming galaxies to the weak lensing signal.

\subsection {The Role of Star-Forming Galaxies}

The role of star-forming galaxies as tracers of the mass distribution has been explored previously by calculating two point correlation functions for quiescent and star-forming galaxies \citep{2002ApJ...571..172Z,2002MNRAS.332..827N,2005ApJ...630....1Z, 2008ApJ...672..153C, 2014ApJ...784..128S,
2015MNRAS.454.2120F}.
 \cite{2002ApJ...571..172Z}, \cite{2002MNRAS.332..827N}, and \cite{2005ApJ...630....1Z} compute two point correlation functions at low redshift. They find that brighter, redder and more massive galaxies are more strongly clustered.
\cite{2008ApJ...672..153C} extend the correlation function analysis to $z\sim 1$. In their analysis of the DEEP2 redshift survey, \cite{2008ApJ...672..153C} show that quiescent galaxies are more strongly clustered than blue objects at this redshift also. In the context of our results, it is interesting that \cite{2008ApJ...672..153C} compute projected two-point correlation functions for the most luminous (and probably most massive) blue galaxies and find a steep rise on small scales possibly reflecting their presence as central galaxies within parent dark matter haloes. However, \cite{2008ApJ...672..153C} also argue that the relation between galaxy color and density  $z \sim$ 1 is comparable with the relation at the current epoch seeming to contradict the increasing lensing signal we observe. Although color selection is not the same as our spectral division of the galaxy population, \cite{2016ApJ...818..174H} find similar results based on spectral classification.

Only a small fraction of galaxies reside in the cores of rich clusters and the weak lensing signal we detect, particularly at the cross-correlation peaks (Figure \ref{fig:corr_z.png}), is most sensitive to these massive systems.
Thus direct observation of the evolution of the galaxy population in clusters is probably more closely related to our results than the two-point correlation function analyses.  
In a search for obscured star-forming members of clusters in the redshift range 
0.2-0.83, \cite {2008ApJ...685L.113S} study the mid-infrared properties of 1315 spectroscopically confirmed cluster members. They find a substantial increase in the fraction of cluster
galaxies with mid-infrared star formation rates $\gtrsim5 M_\odot$ yr$^{-1}$ from 3\% at $z = 0.02$ to 13\% at $z = 0.83$. Inclusion of optically identified star-forming galaxies increases the star forming fraction at $z = 0.83$ to $\sim 23$\%.  For comparison with our lensing results, an interesting conclusion of \cite {2008ApJ...685L.113S} is that the increase in the fraction of star-forming galaxies persists for the most massive cluster members with masses $\gtrsim 10^{10.5} M_\odot$. This stellar mass range overlaps the range that contributes most strongly to our predicted maps. 

\citet{2013ApJ...779..138B} extend 
Spitzer 24 $\mu$ surveys of cluster members to $z \sim 1$. They show that at $z > 1$ clusters have substantial star formation activity at all radii including the cluster core. 
\cite {2008MNRAS.391.1758K} found similar behavior for 15$\mu$ detected sources in and around a cluster at $z\sim0.81$.
Figure \ref{fig:propertiesForHighRedshift} appears to show similar behavior for massive star-forming galaxies  in the
region around the massive X-ray clusters in our survey at $z\sim 0.53$. 
These and other observations underscore  the changing role of massive star-forming galaxies in clusters  with cosmic time. Their role becomes increasingly significant in the higher redshift universe. In contrast, our lensing observations reveal no difference in the signal from the star-forming and quiescent populations in ranges dominated by lower mass groups. This result may reflect the previously observed dependence of the evolution of 
star-formation activity in massive galaxies on the mass of the system where they reside.
For example, \citet{2014MNRAS.445.2725E} show that star-formation activity decreases more
rapidly toward low redshift in more massive systems of galaxies. It is also possible that
our result merely reflects the relative insensitivity of the lensing signal to differences in the distribution of objects around less massive, more poorly populated systems.

In lensing investigations, the star-forming population has largely been ignored.
However, recently \citet{2014MNRAS.437.2111V} used the CFHTLenS weak lensing survey to  evaluate the stellar-to-halo mass of both quiescent and star-forming galaxies.
The typical stellar mass to halo mass ratio is a factor of 0.8--2 greater for quiescent galaxies with $10.5\lesssim\log(M_{\star}/M_{\odot})\lesssim11.5$.

When we compute the normalized zero-lag cross-correlation amplitude for the entire predicted map, we effectively assume a single stellar mass to halo mass ratio for all of the galaxies.
In this approach we obviously underestimate the contribution of quiescent objects and overestimate the contribution of star-forming objects.
One of the strengths of the cross-correlation approach is that when we segregate the populations, the relative amplitudes of the zero-lag cross-correlation are independent of the average difference in stellar mass to halo mass ratio.
Thus the increase with redshift of the zero-lag cross-correlation amplitude for star-forming relative to quiescent galaxies is independent of differences in the stellar mass to halo mass ratio.

The weak lensing maps we explore (Section \ref{starforming}) thus show another view of the evolution of star-forming galaxies in the universe.
This picture of the way they trace the evolving matter distribution in the universe complements the view provided by correlation function analysis and detailed observations of rich clusters. The signal we observe may be dominated by the changing population of galaxies
in and around massive clusters.
Although the comparison we make here extends only to $z \sim 0.6$, Figure \ref{fig:corr_z.png} suggests
a continuing increase in the star-forming galaxy contribution to the cross-correlation signal at even greater redshift. Lensing maps derived from HSC data have the potential to detect structure at  $z\sim$ 1. One of the potentially important future applications of these
maps is the calibration of both quiescent and star-forming galaxies as tracers of the matter distribution.

\section{Conclusion}\label{sec:conclusion}

We use an HSC weak lensing map and a predicted map derived from a dense redshift survey as cosmographic tools to explore the way galaxies trace the matter distribution in the universe. The resolution of the lensing map and the density and completeness of the redshift survey enable separation of the lensing signal from massive quiescent and star-forming galaxies. This separation provides a view of the evolution of the mass content of star-forming galaxies as a function of redshift that complements other approaches to
studying the down-sizing of star-forming galaxies.

The  Subaru/HSC $i$-band imaging of DLS F2 field provides a weak lensing ($\kappa$) map
based on sources with an average surface number density of $\sim$30 arcmin$^2$. The resulting map
has an effective resolution of $\sim 1$ arcmin.

We  construct a predicted  lensing map from a complete spectroscopic survey of the F2 field including 12,705 galaxies with  $R \leq20.6$.
We use the known stellar masses of the individual galaxies as a proxy for the total halo mass;
in effect we construct a total predicted map by assuming an average halo to stellar mass ratio.
The primary contribution to the predicted map comes from galaxies in the stellar mass range $\sim10^{11}M_{\odot}$ where the
galaxy-galaxy lensing measurements of \citet{2014MNRAS.437.2111V} suggest that our assumption of an essentially constant halo mass to stellar mass ratio is reasonable.
We weight the map to account for the unobserved low mass portion of the galaxy stellar mass function.

Cross-correlation is a powerful technique for assessing the relationship between the structure in the $\kappa$ and predicted maps.
We first  cross-correlate the total predicted map with the $\kappa$ map
and detect the zero-lag normalized cross-correlation signal at the $30\sigma$ level. 
The width of the normalized cross-correlation exceeds the smoothing scale and is similar to the scale of a typical massive galaxy cluster at  $z\sim0.3$.
The width of the cross-correlation peak thus suggests that the signal originates largely from massive systems of galaxies.

We explore the impact of structure as a function of redshift by binning the redshift survey into slices with a width of $dz=0.05$.
The zero-lag normalized cross-correlation has significant local maxima at redshifts coinciding with those of  known massive clusters detected independently as X-ray sources.
Even in redshift slices where there are no known massive galaxy clusters, there is still significant signal in the cross-correlation at $\gtrsim3\sigma$ level. This residual signal originates from lower mass groups that trace the large-scale of the universe.

The spectroscopy enables identification of samples of  quiescent galaxies and star-forming galaxies based on the widely used spectral indicator $D_n4000$. We construct predicted maps based on these samples.
The normalized cross-correlation signal from quiescent galaxies is consistent with the signal for the total sample. This conclusion supports previous studies that use red galaxies only to interpret weak lensing maps. 

The  cross-correlation between the $\kappa$ map and predicted maps derived for the star-forming population reveals both the increasing fraction of massive galaxies that are star-forming galaxies and the increasing presence of these massive star-forming galaxies within clusters at greater redshift. The significance of the zero-lag amplitude of the star-forming cross-correlation increases with redshift from $\sim5\sigma$ at $z= 0.3$ to $\sim7\sigma$ at $z=0.5$.  The predicted maps provide striking 
qualitative confirmation of the behavior of the cross-correlation.

The comparison between the $\kappa$ and predicted maps is limited to redshifts $\lesssim 0.6$
by the depth of the redshift survey. The weak lensing map probably contains signal from structure to $z \sim 1$. Deeper redshift surveys combined with similar weak lensing maps 
should reveal the contribution of star-forming galaxies as tracers of the matter distribution in this higher redshift range where star-forming galaxies begin to dominate the population in massive galaxy clusters.

\acknowledgements
We thank the anonymous referee for careful reading of the manuscript and insightful suggestions.
We are very grateful to all of the Subaru Telescope staff.
We thank Dr. Furusawa for providing the Suprime-Cam number counts we use to check our data. 
We thank Dr. Hamana for providing his reconstruction code.
We thank Dr. Oguri for incisive comments that have led to improvements in the analysis and discussion.
We thank Dr. Okabe for sharing computer resources necessary for reduction and analysis.
We thank Dr. Barnacka for a careful reading of the manuscript.

YU was supported by a Grant-in-Aid for Young Scientists (B) from the JSPS (26800103) and MEXT Grant-in-Aid for Scientific Research on Innovative Areas “New Developments in Astrophysics Through Multi-Messenger Observations of Gravitational Wave Sources” (24103003). The Smithsonian Institution supports the research of MJG. A Smithsonian Clay Postdoctoral Fellowship generously supports the research of HJZ.

The Hyper Suprime-Cam (HSC) collaboration includes the astronomical
communities of Japan and Taiwan, and Princeton University.  The HSC
instrumentation and software were developed by the National
Astronomical Observatory of Japan (NAOJ), the Kavli Institute for the
Physics and Mathematics of the Universe (Kavli IPMU), the University
of Tokyo, the High Energy Accelerator Research Organization (KEK), the
Academia Sinica Institute for Astronomy and Astrophysics in Taiwan
(ASIAA), and Princeton University.  Funding was contributed by the
Ministry of Education, Culture, Sports, Science and Technology (MEXT),
the Japan Society for the Promotion of Science (JSPS), 
(Japan Science and Technology Agency (JST),  the Toray Science 
Foundation, NAOJ, Kavli IPMU, KEK, ASIAA,  and Princeton University.

We used software developed for the Large Synoptic Survey Telescope. We thank the LSST Project for making their code available as free software at http://dm.lsstcorp.org \citep{2008arXiv0805.2366I,2010SPIE.7740E..15A}.

We acknowledge use of {\it lensfit} for shear measurement \citep{2007MNRAS.382..315M,2008MNRAS.390..149K}.

Funding for SDSS-III has been provided by the Alfred P. Sloan Foundation, the Participating Institutions, the National Science Foundation, and the U.S. Department of Energy Office of Science. The SDSS-III web site is http://www.sdss3.org/.

SDSS-III is managed by the Astrophysical Research Consortium for the Participating Institutions of the SDSS-III Collaboration including the University of Arizona, the Brazilian Participation Group, Brookhaven National Laboratory, Carnegie Mellon University, University of Florida, the French Participation Group, the German Participation Group, Harvard University, the Instituto de Astrofisica de Canarias, the Michigan State/Notre Dame/JINA Participation Group, Johns Hopkins University, Lawrence Berkeley National Laboratory, Max Planck Institute for Astrophysics, Max Planck Institute for Extraterrestrial Physics, New Mexico State University, New York University, Ohio State University, Pennsylvania State University, University of Portsmouth, Princeton University, the Spanish Participation Group, University of Tokyo, University of Utah, Vanderbilt University, University of Virginia, University of Washington, and Yale University.

The authors wish to recognize and acknowledge the very significant cultural role and reverence that the summit of Mauna Kea has always had within the indigenous Hawaiian community. We are most fortunate to have the opportunity to conduct observations from this sacred mountain.

\facility{Subaru (HSC), MMT (Hectospec)}

\bibliographystyle{aasjournal}
\bibliography{crosscorrelation}

\end{document}